\documentclass[showpacs,aps,prd,nofootinbib,floatfix,amsmath,amssymb]{revtex4}
\usepackage{graphicx}
\usepackage[tight,TABTOPCAP]{subfigure}
\usepackage{color}
\begin{document}

\makeatletter
%Feynman slash
\newbox\slashbox \setbox\slashbox=\hbox{$/$}
\newbox\Slashbox \setbox\Slashbox=\hbox{\large$/$}
\def\pFMslash#1{\setbox\@tempboxa=\hbox{$#1$}
  \@tempdima=0.5\wd\slashbox \advance\@tempdima 0.5\wd\@tempboxa
  \copy\slashbox \kern-\@tempdima \box\@tempboxa}
\def\pFMSlash#1{\setbox\@tempboxa=\hbox{$#1$}
  \@tempdima=0.5\wd\Slashbox \advance\@tempdima 0.5\wd\@tempboxa
  \copy\Slashbox \kern-\@tempdima \box\@tempboxa}
\def\FMslash{\protect\pFMslash}
\def\FMSlash{\protect\pFMSlash}
\def\miss#1{\ifmmode{/\mkern-11mu #1}\else{${/\mkern-11mu #1}$}\fi}
%%%% Uso:  \pFMSlash{p}
\makeatother

%\tightenlines
\title{Effects of Lorentz violation through the $\gamma e\to W\nu_e$ process in the Standard Model Extension}
\author{J. I. Aranda$^{(a)}$, F. Ram\'\i rez-Zavaleta$^{(a)}$, D. A. Rosete$^{(b)}$, F. J. Tlachino$^{(b)}$, J. J. Toscano$^{(b)}$, E. S. Tututi$^{(a)}$}
\address{
$^{(a)}$Facultad de Ciencias F\'\i sico Matem\' aticas,
Universidad Michoacana de San Nicol\' as de
Hidalgo, Avenida Francisco J. M\' ujica S/N, 58060, Morelia, Michoac\'an, M\' exico. \\
$^{(b)}$Facultad de Ciencias F\'{\i}sico Matem\'aticas,
Benem\'erita Universidad Aut\'onoma de Puebla, Apartado Postal
1152, Puebla, Puebla, M\'exico.}
\begin{abstract}
Physics beyond the Fermi scale could show up through deviations of the gauge couplings predicted by the electroweak Yang-Mills sector. This possibility is explored in the context of the International Linear Collider (ILC) through the helicity amplitudes for the $\gamma e\to W\nu_e$ reaction to which contributes the trilinear $WW\gamma$ coupling. The new physics effects on this vertex are parametrized in a model-independent fashion through an effective electroweak Yang-Mills sector, which is constructed by considering two essentially different sources of new physics. In one scenario, Lorentz violation will be considered exclusively as the source of new physics effects. This type of new physics is considered in an extension of the Standard Model that is known as the Standard Model Extension (SME), which is an effective field theory that contemplates $CPT$ and Lorentz violation in a model-independent fashion. Any source of new physics that respects the Lorentz symmetry, will be considered within the general context of the well known Conventional Effective Standard Model (CESM) extension. Both the SME and the CESM descriptions include gauge invariant operators of dimension higher than four, which, in general, transform as Lorentz tensors of rank higher than zero. Whereas in the former theory observer Lorentz invariants are constructed by contracting these operators with constant Lorentz tensors, in the latter the corresponding Lorentz invariant interactions are obtained contracting such operators with products of the metric tensor. In this work, we focus on a dimension six Lorentz 2-tensor, ${\cal O}_{\alpha \beta}$, which arises from an effective $SU(2)_L$ Yang-Mills sector. When this operator is contracted with a constant antisymmetric background tensor, $b^{\alpha \beta}$, the corresponding observer invariant belongs to the SME, whereas if it is contracted with the metric tensor, $g_{\alpha \beta}$, an effective interaction in the context of the CESM is obtained. We focus our study on the possibility of experimentally distinguish both types of new physics effects on the $WW\gamma$ vertex. It is found that for a new physics scale of the same order of magnitude and under determined circumstances, both types of new physics effects will be clearly distinguished.
\end{abstract}


\maketitle
\section{Introduction}
\label{i}There are founded reasons to think that the Lorentz symmetry can be violated at very small distances or very high energies. For instance, it has been discovered that certain mechanism in string theory~\cite{KS} or within the context of quantum gravity~\cite{QG} can cause the violation of Lorentz symmetry. Since these theories have not been sufficiently developed, an effective field theory that contain both the Standard Model (SM) and gravity has been formulated. This effective theory, for which exists a minimal version without gravity~\cite{SME}, is called the Standard Model Extension (SME)~\cite{SMEG}. Although motivated from specific scenarios in the context of string theory or general relativity with spontaneous symmetry breaking, the SME is beyond these specific ideas due to its generality, which is the main advantage of effective field theories. Thus the SME provides us a powerful tool for investigating Lorentz violation in a model-independent manner. Lorentz violation is also a feature of field theories formulated in a noncommutative space-time~\cite{Snyder}. This type of theories have been the subject of interest since Seiberg and Witten showed how to connect commutative and noncommutative gauge theories~\cite{SWM}. A method to formulate the SM as an effective theory (NCSM), which is expressed in powers of the noncommutativity parameter, has been proposed~\cite{NCYM,NCSM}. Indeed, as it has been shown in Ref.~\cite{SME-NCSM}, the NCSM is a subset of the SME. Although these effective theories introduce constant background fields that carry Lorentz indices, they are not Lorentz invariants under general Lorentz transformations, but only under observer Lorentz transformations. As it has been discussed in references~\cite{SME-NCSM}, there are two distinct classes of Lorentz transformations, namely, the observer and particle Lorentz transformations. The former class of Lorentz transformations corresponds to a change of coordinates, whereas the latter can be associated with a change of the measurement apparatus~\cite{Rev}.

In this paper, we are interested in investigating the $\gamma e\to W\nu_e$ processes in presence of a constant background field characterized by an antisymmetric tensor $b^{\alpha \beta}$, which can arise from quantum gravity with spontaneous symmetry breaking or from a noncommutative spacetime. This is an interesting reaction, which will be under the scrutiny of The International Linear Collider (ILC)~\cite{ILC} operating in the $e\gamma$ mode. This ambitious program of electron-positron collisions at the TeV scale, will provide a clean environment to make studies beyond the capabilities of the Large Hadron Collider(LHC), which will expand our knowledge of the standard model and promise access to new physics that could eventually show up at this energy scale. One interesting feature of this reaction is that the photon is on-shell, in contrast with $e^+e^-$ collisions, which occur through highly off-shell photons (and also through $Z$ gauge bosons). Although the radiative corrections are important within the SM, for our purposes it is sufficient with comparing  our results with the SM prediction at the Born approximation. In the SM, the tree-level cross section for $\gamma e\to W\nu_e$ has already been computed a long time ago~\cite{TSM}. A comprehensive analysis of the one-loop radiative corrections was given in Ref.~\cite{RSM}. This process provides a good mechanism to investigate the presence of new physics effects on the $WW\gamma$ vertex with independence of the $WWZ$ one and thus would be complementary of the $e^+e^-\to W^-W^+$ reaction, to which contribute simultaneously both $WW\gamma$ and $WWZ$ vertices. New physics effects on the $\gamma e\to W\nu_e$ reaction has been studied by some authors~\cite{W1,W2,W3,W4,W5,W6} in the context of effective field theories~\cite{EL}. Such a new physics has been considered through an effective vertex $W^-W^+\gamma$, which is parametrized by means of the form factors that characterize the electromagnetic properties of the $W$ gauge boson. To our best knowledge, Gaemers and Gounaris~\cite{GG} derived initially 9 form factors for the $WW\gamma$ vertex, but further on a careful analysis carried out by Hagiwara-Peccei-Zeppenfeld-Hikasa~\cite{Hagiwara} showed that only 7 of these quantities are independent indeed.  These form factors define the charge, the magnetic and electric dipole moments, the magnetic and electric quadrupole moments, and the CP-even and CP-odd anapole moments of this particle. Although model independent, it is assumed that these form factors respect both the Lorentz symmetry and the SM gauge symmetry. In other words, the sources of new physics have nothing to do with Lorentz violation. As already mentioned, in this work we are interested in researching deviations of the SM prediction for the $\gamma e\to W\nu_e$ process by assuming the presence of a Lorentz violating effective $WW\gamma$ vertex, whose source may be, for instance, general relativity with spontaneous symmetry breaking or a noncommutative spacetime. However, we will adopt a model-independent approach by using the general formalism of the SME~\cite{SME}. The structure of the effective Lagrangian characterizing the SME~\cite{SME,T1} differs substantially from that describing the Conventional Effective Standard Model (CESM)~\cite{EL} extension. While the SME is constructed out by gauge-invariant Lorentz tensors contracted with constants Lorentz tensors specifying preferred spatial directions, the CESM are made of objects that are both gauge-invariant and Lorentz-invariant or, equivalently, of gauge-invariant Lorentz tensors appropriately contracted with products of metric tensors. Due to this, it is expected that the SM deviations induced by an anomalous $WW\gamma$ vertex on the $\gamma e\to W\nu_e$ process differs from one to other approach. An important goal of this work is to investigate not only the deviations of the SM prediction to the  $\gamma e\to W\nu_e$ reaction due to Lorentz violating effects present in the $WW\gamma$ vertex, but also to compare these deviations with those induced by other sources of new physics effects parametrized in the scheme of CESM. This type of information will be valuable in future experiments. We will focus on the Yang-Mills part of the effective Lagrangian that characterizes the SME (or also the NCSM) modified by the presence of an observer invariant that arises from the contraction of the constant antisymmetric tensor  $b^{\alpha \beta}$ with a Lorentz 2-tensor that is invariant under the $SU_L(2)$ gauge group.  This extended Yang-Mills sector generates a nonrenormalizable $WW\gamma$ vertex, which differs substantially from the one studied in references~\cite{W1,W2,W3,W4,W5,W6} within the context of the CESM~\cite{EL}. In general, for each observer-invariant constructed with a Lorentz $k$-tensor contracted with a $k$-tensor background field in the SME, there is a counterpart in the context of the CESM that result from the contraction of such Lorentz $k$-tensor with an appropriately product of the metric tensor. To simplify our analysis as much as possible, we will consider the simplest extension of the $SU_L(2)$ Yang-Mills sector in both the SME and the CESM. Explicit expressions for the helicity amplitudes of the $\gamma e\to W\nu_e$ reaction including a comprehensive analysis of their angular distributions will be presented.

The rest of the paper has been organized as follows. In Sec. \ref{L}, effective Lagrangians for the Yang-Mills sector of the $SU_L(2)$ group that includes gauge-invariant interactions of up to dimension-six in both the CESM and the SME are presented. In particular, the main differences of the gauge and Lorentz structure of the $WW\gamma$ vertex arising from each of these effective formulations of new physics is discussed. Sec. \ref{ha} is devoted to calculate the helicity amplitudes for the $\gamma e\to W\nu_e$ reaction. In Sec. \ref{D}, we discuss our results. Finally, in Sec. \ref{C}, the conclusions are presented.

\section{Effective Yang-Mills Lagrangian }
\label{L}The gauge structure of the $WW\gamma$ vertex (and also of $WWZ$) has been the subject of important attention in the literature in diverse contexts. The one-loop radiative corrections to the renormalizable vertex have been calculated in the SM~\cite{Bardeen} and some of its extensions~\cite{BSM}. The radiative corrections to these vertices with the $\gamma$ and $Z$ bosons off shell have been studied in the SM using a linear gauge~\cite{SMLG} and also via the Pinch Technique~\cite{PT}. Virtual effects of new heavy gauge bosons to these off shell vertices have been studied in a covariant way under the electroweak group within the context of $331$ models~\cite{T2} and in theories with universal extra dimensions~\cite{ED}. Its most general structure has been parametrized in  model independent manner in the context of CESM~\cite{GG,Hagiwara} and used in countless phenomenological applications~\cite{W1,W2,W3,W4,W5,W6,Other}. As commented, the $WW\gamma$ effective vertex that arises from the CESM approach differs substantially from the one that can be constructed in the context of the SME. To clarify the main idea behind our work, let us to discuss with some extent how the $WW\gamma$ vertex emerges in both the CESM and SME descriptions of new physics.

The building blocks needed to introduce $SU_L(2)$ and $U_Y(1)$ invariant operators of arbitrary dimension are the respective curvatures $W_{\mu \nu}=T^aW^a_{\mu \nu}$ and $B_{\mu \nu}$. These gauge invariant operators are all Lorentz tensors of even rank, which will be denoted by the moment as ${\cal O}_{\mu_1,\mu_2,\cdots \mu_{2n}}$. One can to construct operators that are invariant under general Lorentz transformations by contracting these gauge invariant operators with an tensor of the same rank made of a product of metric tensors, that is, $g^{\mu_1,\mu_2}\cdots g^{\mu_{2n-1}\mu_{2n}}{\cal O}_{\mu_1,\mu_2,\cdots \mu_{2n}}$ . Alternatively, one can to construct Lorentz observer invariant operators by contracting these gauge invariant operators with a constant tensor of the same rank, that is,  $b^{\mu_1 \mu_2\cdots \mu_{2n}}{\cal O}_{\mu_1,\mu_2,\cdots \mu_{2n}}$. The former scheme leads us to the CESM, which is a technique that allows us to parametrize, in a model-independent fashion, effects of new physics that respect both the gauge symmetry and the Lorentz symmetry. On the other hand, when one adopts the latter scheme, one arrives to the SME~\cite{SME}, which is an effective field theory that allows to incorporate $CPT$ violation and Lorentz violation in a model-independent manner. It should be noticed that the gauge structure is the same in both CESM and SME approach to new physics. The fundamental difference between both schemes arises from the way through which the Lorentz invariant action is constructed. In the CESM approach, Lorentz invariance is established through contractions with the metric tensor, which is a self-invariant Lorentz object by definition. In contrast, in the SME approach to new physics, a Lorentz observer invariant action can be constructed by contracting the Lorentz $2n$-tensor operators with constant background $2n$-tensors which are true tensorial objects under observer Lorentz transformations, but not under particle Lorentz transformations~\cite{SME,Rev}. This general scheme comprises the very interesting situation in which the $b^{\mu_1 \mu_2\cdots \mu_{2n}}$ constant tensor corresponds to a vacuum expectation value of a tensor field $B^{\mu_1 \mu_2\cdots \mu_{2n}}(x)$. This particular case corresponds to a spontaneous symmetry breaking of the Lorentz group, which arises in specific scenarios of string theories of general relativity.

\begin{figure}
\centering\includegraphics[width=1.5in]{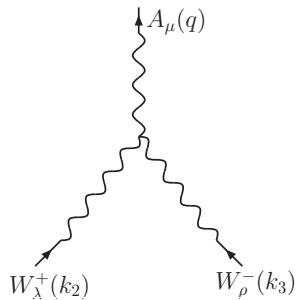}
\caption{\label{V}The trilinear $WW\gamma$ vertex.}
\end{figure}

As commented in the introduction, in this work we will focus on the electroweak Yang-Mills sector of the SM. The gauge invariant Lorentz tensor operators of up to dimension six that can be constructed with the $W_{\mu \nu}$ and $B_{\mu \nu}$ curvatures are the following\footnote{Other possible dimension-six Lorentz 2-tensor that can be constructed is $Tr[W_{\alpha \beta}W_{\mu \nu}W^{\mu \nu}]$, but it vanishes, as $ W^b_{\mu \nu}W^{c\mu \nu}Tr[\sigma^a \sigma^b \sigma^c]=2i W^b_{\mu \nu}W^{c\mu \nu}\epsilon^{abc}=0$.}:
\begin{eqnarray}
SU_L(2): &&  {\cal O}^W_{\alpha \beta \lambda \rho}=Tr[W_{\alpha \beta}W_{\lambda \rho}]\, , \, \, {\cal O}_{\alpha \beta}=Tr[W_{\alpha \lambda} W_{\beta \rho} W^{\lambda  \rho}] \, , \\
U_Y(1): && {\cal O}^B_{\alpha \beta \lambda \rho}=B_{\alpha \beta}B_{\lambda \rho}\, .
\end{eqnarray}
In the context of the CESM extension, the contractions $g^{\alpha \beta}g^{\lambda \rho}{\cal O}^W_{\alpha \beta \lambda \rho}$, $g^{\alpha \beta}g^{\lambda \rho}{\cal O}^B_{\alpha \beta \lambda \rho}$, and $g^{\alpha \beta}{\cal O}_{\alpha \beta}$ lead to an effective electroweak Yang-Mills sector that includes up to dimension six interactions, which can conveniently be written as
\begin{equation}
{\cal L}_{CESM}^{YM}=-\frac{1}{4}W^a_{\mu \nu}W^{\mu \nu}_a-\frac{1}{4}B_{\mu \nu}B^{\mu \nu}+\frac{g\, \alpha_W}{\Lambda^2}\frac{\epsilon_{abc}}{3!}W^a_{\alpha \lambda}W^{b \alpha}_{\, \, \, \rho}W^{c\lambda \rho}\, ,
\end{equation}
where some constant factors have been introduced. In particular, $\Lambda$ represents the new physics scale and $\alpha_W$ is an unknown coefficient, which can be calculated once the fundamental theory is known. To write the most general Lorentz structure of the $WW\gamma$ vertex it is necessary to introduce additional dimension-six operators whose construction involves the Higgs doublet~\cite{EL,GG,Hagiwara}, but we do not consider they here. It should be noticed that due to the symmetric character of the metric tensor, only the symmetric part of the ${\cal O}_{\alpha \beta}$ operator contributes to the $WW\gamma$ coupling. In this context, the anomalous part of the $WW\gamma$ vertex\footnote{The SM structure of the $WW\gamma$ vertex is well known and we do not present here.} is given by the Lagrangian
\begin{equation}
{\cal L}^{CESM}_{WW\gamma}=\frac{ie\alpha_W}{\Lambda^2}W^{-}_{\lambda \rho}W^{+\lambda}_{\, \, \eta}F^{\rho \eta}
\end{equation}
and the corresponding vertex function can be written as
\begin{equation}
\Gamma^{CESM}_{\lambda \rho \mu}(q,k_1,k_2)=\frac{ie\alpha_W}{\Lambda^2}\left(q^\eta\delta^\beta_\mu-q^\beta \delta^\eta_\mu \right)\left(k^\alpha_2g_{\eta \lambda}-k_{2\eta}\delta^\alpha_\lambda \right)\left(k_{3\alpha}g_{\beta \rho}-k_{3\beta}g_{\alpha \rho} \right)\, ,
\end{equation}
where the notation and conventions shown in Fig. \ref{V} were used. Notice that this vertex satisfies the following simple Ward identities
\begin{eqnarray}
q^\mu \Gamma^{CEFT}_{\lambda \rho \mu}(q,k_1,k_2)&=&0\, ,\\
k^\lambda_2\Gamma^{CEFT}_{\lambda \rho \mu}(q,k_1,k_2)&=&0\, ,\\
k^\rho _3\Gamma^{CEFT}_{\lambda \rho \mu}(q,k_1,k_2)&=&0\, .
\end{eqnarray}

Having discussed the structure of the $WW\gamma$ vertex within the context of CESM, we proceed to present an effective Yang-Mills Lagrangian that generates this vertex in the context of the SME. The main differences between the CESM and the SME, as well as the fact that the NCSM is a subset of the SME~\cite{SME-NCSM}, have been discussed with some extent in reference~\cite{T1}. Here, we will present only those features that are relevant for our discussion. As already commented, we will consider a minimal scenario in which not degrees of freedom different of the gauge fields associated with the electroweak group are considered. In addition, our scenario will be one in which constant background tensors couples with the gauge curvatures, but not with their dual. Within the context of the SME, an observer Lorentz invariant and $CPT$ conserving effective electroweak Yang-Mills sector can be constructed by contracting the ${\cal O}^W_{\alpha \beta \lambda \rho}$, ${\cal O}^B_{\alpha \beta \lambda \rho}$, and ${\cal O}_{\alpha \beta}$ operators with constant Lorentz tensors. The corresponding Lagrangian can be written as follows:
\begin{equation}
{\cal L}^{YM}_{SME}=-\frac{1}{4}W^a_{\mu \nu}W^{\mu \nu}_a-\frac{1}{4}B_{\mu \nu}B^{\mu \nu}+k^{\alpha \beta \lambda \rho}_W{\cal O}^W_{\alpha \beta \lambda \rho}+k^{\alpha \beta \lambda \rho}_B{\cal O}^B_{\alpha \beta \lambda \rho}+b^{\alpha \beta}{\cal O}_{\alpha \beta} \, ,
\end{equation}
where the SM Yang-Mills sector has been included. The dimensionless constant tensors $k^{\alpha \beta \lambda \rho}_{W,B}$ are antisymmetric under the interchanges $\alpha \leftrightarrow \beta$ and $\lambda \leftrightarrow \rho$, but are symmetric under the simultaneous interchange of the pairs of indices $(\alpha \beta) \leftrightarrow (\lambda \rho)$~\cite{SME}. On the other hand, it is assumed that the constant 2-tensor $b^{\alpha \beta}$, which has units of mass squared, is antisymmetric. This means that only the antisymmetric part of the ${\cal O}_{\alpha \beta}$ operator contributes to the SME, in contrast with the CESM approach, in which only the symmetric part of this operator contributes. This is a crucial feature that allows us to distinguish one approach from the other. In particular, as we will see below, the symmetric or antisymmetric anomalous contributions of ${\cal O}_{\alpha \beta}$ to the $WW\gamma$ vertex will be reflected in the helicity amplitudes for the $\gamma e \to W\nu_e$ process. As discussed in the context of string theory quantization~\cite{SWM} and in general relativity with spontaneous symmetry breaking~\cite{SSBQG}, there exists more than a simple analogy between the six $b^{\alpha \beta}$ quantities and the six components of the electromagnetic field tensor $F^{\alpha \beta}$. These six independent components, given by $e^i\equiv \Lambda^2_{LV} b^{0i}$ and $b^i\equiv (1/2)\Lambda^2_{LV}\epsilon^{ijk}b^{jk}$, with $\Lambda_{LV}$ the new physics scale, determine two preferred spatial directions, which play the role of an external agent that would induce deviations from the SM predictions which in principle could be observed in future high-energy experiments.

As far as the renormalizable operators ${\cal O}^W_{\alpha \beta \lambda \rho}$ and  ${\cal O}^B_{\alpha \beta \lambda \rho}$ are concerned, they have already been considered in the literature in other contexts. Besides to modify the $WW\gamma$ vertex, these operators also introduce changes in the photon propagator and through it induce contributions to some cosmological observables, which impose severe constraints on these class of operators~\cite{KM,KR}. Due to this, in this work we will not consider Lorentz violating effects arising from these renormalizable interactions. Then, the anomalous contribution to the $WW\gamma$ vertex in the context of the SME arises only from the nonrenormalizable term $b^{\alpha \beta}{\cal O}_{\alpha \beta}$. This contribution is given by
\begin{equation}
\mathcal{L}^{NC}_{WW\gamma}= \frac{ie}{2}b^{\alpha
\beta}(W^-_{\alpha \lambda}W^+_{\beta \rho}F^{\lambda
\rho}+W^+_{\alpha \lambda}W^{-\lambda \rho}F_{\beta
\rho}+W^-_{\beta \rho}W^{+\lambda \rho}F_{\alpha \lambda}).
\end{equation}
Using the notation and conventions shown in Fig.\ref{V}, the corresponding vertex function can be written as follows:
\begin{equation}
\Gamma_{\mu \lambda \rho}(q,k_2,k_3)=\frac{ie}{2}\,b^{\alpha \beta}T^{\eta \xi}_{\mu}\Gamma_{\alpha \beta \eta \xi \lambda \rho}\, \, ,
\end{equation}
where
\begin{equation}
T^{\eta \xi}_{\mu}=q^{\xi}\delta^{\eta}_{\mu}-q^{\eta}\delta^{\xi}_{\mu}
\end{equation}
and
\begin{eqnarray}
\Gamma_{\alpha \beta \eta \xi \lambda \rho}(k_2,k_3) & = & +(k_{2 \beta}g_{\xi \lambda}-k_{2 \xi}g_{\beta \lambda})(k_{3 \alpha}g_{\eta \rho}-k_{3 \eta}g_{\alpha \rho}) {}\nonumber \\
&& {} +g_{\eta \beta}(k_{2 \alpha}g_{\sigma \lambda}-k_{2 \sigma}g_{\alpha \lambda})(k_3^{\sigma}g_{\xi \rho}-k_{3 \xi}\delta^{\sigma}_{\rho}) {}\nonumber \\
&& {} +g_{\eta \alpha}(k_{2 \xi}\delta^{\sigma}_{\lambda}-k_{2 \sigma}g_{\xi \lambda})(k_{3 \beta}g_{\sigma \rho}-k_{3 \sigma}g_{\beta \rho}) \, \, .
\end{eqnarray}
From this expression, it is evident that $\Gamma_{
\mu\lambda \rho}(q,k_2,k_3)$  satisfies the following simple Ward
identities:
\begin{eqnarray}
q^\mu \Gamma_{\mu\lambda \rho}(q,k_2,k_3)=0, \\
k^\lambda_2\Gamma_{\mu\lambda \rho}(q,k_2,k_3)=0, \\
k^\rho_3\Gamma_{\mu\lambda \rho}(q,k_2,k_3)=0.
\end{eqnarray}
The first of these identities guaranties electromagnetic gauge invariance of the $WW\gamma$ anomalous contribution to the $e\gamma \to W\nu_e$ process, whereas the last two show that the longitudinal component of the $W$ propagator does not contribute to this process.

\section{The $\gamma e\to W\nu_e$ process} \label{ha}
We now turn to calculate the helicity amplitudes for the $\gamma e\to W\nu_e$ process in the context of an anomalous $WW\gamma$ vertex that arises in both the CESM and the SME. We will present our results in the center of mass reference frame.

\subsection{Kinematics}
Our notation and conventions for the kinematics involved in the process
are shown in Figs.\ref{FD} and \ref{KandBF}, where the photon momentum is taken
along the $+z$ direction. The Lorentz indices and momenta
are specified as follows:
\begin{equation}
A_\mu(k_1)+e^-(p_1)\to W^-_\nu(k_2)+\nu_e(p_2).
\end{equation}
Then, in this frame, the momenta and polarization vectors can be
written as follows:
\begin{eqnarray}
p^\mu_1&=&\frac{\sqrt{s}}{2}(1,0,0,1), \\
k^\mu_1&=&\frac{\sqrt{s}}{2}(1,0,0,-1), \\
p^\mu_2&=&\frac{s-m^2_W}{2\sqrt{s}}(1,\sin\theta,0,\cos\theta),\\
k^\mu_2&=&\frac{s+m^2_W}{2\sqrt{s}}(1,-r \sin\theta,0,-r\cos\theta),
\end{eqnarray}

\begin{eqnarray}
&&\epsilon^\mu(k_1,\lambda_\gamma)=\frac{1}{\sqrt{2}}(0,1,-i\lambda_\gamma,0),
\\
&&\epsilon^{*\nu}(k_2,\lambda_W)=\frac{1}{\sqrt{2}}(0,\cos\theta,i\lambda_W,-\sin\theta),
\\
&&\epsilon^{*\nu}(k_2,0)=\frac{s+m^2_W}{2m_W\sqrt{s}}(r,-\sin\theta,0,-\cos\theta),
\end{eqnarray}
where
\begin{equation}
r=\frac{s-m^2_W}{s+m^2_W}.
\end{equation}
Here, $\lambda_W$ stands for the transverse polarization
states of the $W$ gauge boson. On the other hand, the Mandelstam
variables are given by
\begin{eqnarray}
s&=&(k_1+p_1)^2,\\
t&=&(k_1-k_2)^2=-(s-m^2_W)\sin^2\frac{\theta}{2}, \\
u&=&(k_1-p_2)^2=-(s-m^2_W)\cos^2\frac{\theta}{2}.
\end{eqnarray}

The polarized differential cross section can be written as
follows:

\begin{align}\label{sediff}
\left(\frac{d\,\sigma_{\tiny{\lambda_\gamma\,\bar{\lambda}_W}}}
{d\Omega}\right)_{\tiny{\mathrm{CM}}}=\frac{1}{64\,\pi^2}\,
\frac{s-m_W^2}{s^2}\,|\mathcal{M}_{\tiny{\lambda_\gamma\,\bar{\lambda}_W}}|^2,
\end{align}
where $\mathcal{M}_{\tiny{\lambda_\gamma\,\bar{\lambda}_W}}$ are
the helicity amplitudes, whose transverse and longitudinal
components are defined by
\begin{equation}
\mathcal{M}_{\tiny{\lambda_\gamma\,\bar{\lambda}_W}}=\left\{\begin{array}{ll}
\mathcal{M}_{\tiny{\lambda_\gamma\,\lambda_W}},&\,\,\, \lambda_W=\pm 1 \\
\mathcal{M}_{\tiny{\lambda_\gamma\,\lambda_W^0}},&\,\,\,\lambda_W^0=0.
\end{array}\right.
\end{equation}
Due to the energies used in the study of the process, we drop the electron mass. In this approximation,  the electron and neutrino are left-handed.

The helicity amplitudes can be written as follows:
\begin{equation}\label{totamp}
{\cal M}_{\lambda_\gamma\, \, \bar{\lambda}_W}={\cal M}^{ SM}_{\lambda_\gamma\, \, \bar{\lambda}_W}+{\cal M}^{ NP}_{\lambda_\gamma\, \, \bar{\lambda}_W}
\end{equation}
where the superscripts $SM$ and $NP$ stand for the SM and new physics contributions, respectively. As already commented, effects of physics beyond the Fermi scale on the $WW\gamma$ vertex will be considered in two essentially different model-independent schemes, namely the CESM extension, which respects both the Lorentz and SM gauge symmetries, and the SME, which respect the SM gauge symmetry but violates the Lorentz one.

\subsection{The Standard Model helicity amplitudes}
\begin{figure}
\centering
\includegraphics[width=4.0in]{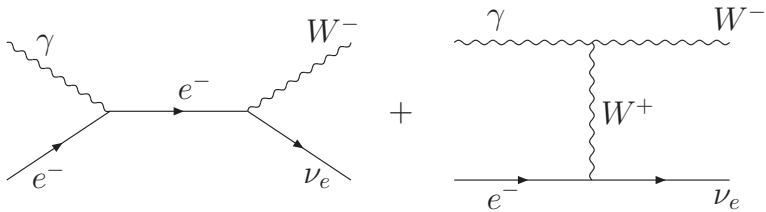}
\caption{\label{FD} Feynman diagrams contributing to the $\gamma e\to W\nu_e$ reaction at the tree level.}
\end{figure}

In the Born approximation, the SM contribution to the $e\gamma \to W\nu_e$ process is given through the Feynman diagrams shown in Fig. \ref{FD}. The corresponding amplitude is given by
 \begin{equation}
\mathcal{M}^{SM}_{\lambda_\gamma\, \, \bar{\lambda}_W} =\frac{e g}{\sqrt{2}} \, \overline u (p_2) \,
P_{R}\Gamma^{SM}_{\mu\nu} \, u(p_1)\, \,
\varepsilon^{\mu}(k_1,\lambda_{\gamma})\,\,\varepsilon^{\ast
\nu}(k_2,\bar{\lambda}_{W}),
\end{equation}
where
\begin{equation}
\Gamma_{\mu\nu}^{SM}=- \frac{i}{s}\, \left\{
 \, \gamma _{\nu }\,(\pFMSlash{p}_1+\pFMSlash{k}_1)\,\gamma _{\mu }
 +\frac{s}{(t-\text{$m_W^{2}$})}
 \left[
2\gamma _{\mu }\,\text{k}_{1\nu }+2\gamma _{\nu }\,\text{k}_{2\mu
}-g_{\mu \nu }\left(\pFMSlash{k}_1+\pFMSlash{k}_2\right)\right] \right\},
\end{equation}
 $P_{R}=\frac{1}{2}(1+\gamma_{5})$ is the right-handed projector,  and we have explicitly used the transversality conditions:  $k_{1}\cdot \varepsilon (k_1,\lambda_\gamma)=0$ and  $k_{2}\cdot \varepsilon^{\ast} (k_2,\bar{\lambda}_{W})=0$. The calculations are carried out in the unitary gauge.  Notice that $k^\mu_{1}\,\overline u (p_2) \,P_{R}\Gamma^{SM}_{\mu\nu} \, u(p_1)=0$, which reflects electromagnetic gauge invariance. The corresponding helicity amplitudes can be written as follows
\begin{eqnarray}\label{SMAMP1}
\mathcal{M}^{\tiny{SM}}_{\tiny{\lambda_\gamma\,\lambda_W}}&=&
-\frac{i\,\sqrt{2}\, \pi \, \alpha}{\, s_W }\,\frac{\sqrt{1-x}\,  \cos
\left(\frac{\theta }{2}\right)}{1+x-(1-x) \cos\theta}
\Big(\lambda_W+ \lambda_\gamma-3\,(1+\, \lambda_\gamma\,
\lambda_W)+x\,
(\lambda_W-1) (\lambda_\gamma -3)\nonumber\\
&&+(x\, (\lambda_W-1)-\lambda_W-1) (\lambda_\gamma +1) \cos \theta \Big),
\end{eqnarray}
\begin{equation}\label{SMAMP2}
\mathcal{M}^{\tiny{SM}}_{\tiny{\lambda_\gamma\,\lambda_W^0}}=
-\frac{ i\,8\, \pi\,\alpha  }{s_W}\, \frac{\sqrt{1-x}\, \sqrt{x}\,
(\lambda_\gamma +1)\, \cos ^2\left(\frac{\theta }{2}\right)\, \sin
\left(\frac{\theta }{2}\right)}{1+x-(1-x) \cos\theta},
\end{equation}
where $x=m_{W}^{2}/s$ and $s_{W}=\sin\theta_{W}$,  being  $\theta_{W} $ the Weinberg angle. Notice that $\mathcal{M}^{\tiny{SM}}_{\tiny{-\,+}}=0$ and $\mathcal{M}^{\tiny{SM}}_{\tiny{-\,0}}=0$ at this order of perturbation theory.

\subsection{New physics effects in the context of CESM}
In this scenario of physics beyond the SM that respects the Lorentz symmetry, only the symmetric part of the ${\cal O}_{\alpha \beta}$ operator contributes to the $e\gamma \to W\nu_e$ reaction. The corresponding contribution to the amplitude can be written as
\begin{equation}
{\cal M}^{CESM}_{\lambda_\gamma \, \, \bar{\lambda}_W}={\cal M}^{SM}_{\lambda_\gamma \, \, \bar{\lambda}_W}-\frac{ieg\alpha_W}{\sqrt{2}\Lambda^2}\frac{1}{t-m^2_W}\left[\bar{u}(p_2)P_R\gamma^\alpha u(p_1)\right]\Gamma^{CESM}_{\alpha \mu \nu}\varepsilon^{\mu}(k_1,\lambda_1)\varepsilon^{*\nu}(k_2,\bar{\lambda}_W)\, ,
\end{equation}
where
\begin{equation}
\Gamma^{CESM}_{\alpha \mu \nu}=\left(k^\rho_1\delta^\eta_\mu-k^\eta_1\delta^\rho_\mu \right)\left(k^\lambda_2g_{\eta\nu}-k_{2\eta}\delta^\lambda_\nu \right)\left[\left(k_2-k_1\right)_\lambda g_{\rho \alpha}-(k_2-k_1)_\rho g_{\lambda \alpha}\right].
\end{equation}
 After some algebra, one obtains for the helicity amplitudes
\begin{equation}
{\cal M}^{CESM}_{\lambda_\gamma \, \, \lambda_W}-{\cal M}^{SM}_{\lambda_\gamma \, \, \lambda_W}=\frac{i2\sqrt{2}\pi \alpha}{s_W}\, \alpha_W\, \left(\frac{s}{\Lambda^2}\right)\frac{\sqrt{1-x}\sin^2\left(\frac{\theta}{2}\right)\cos\left(\frac{\theta}{2}\right)}{1+x-(1-x)\cos\theta }\left[1-\lambda_\gamma \lambda_W-(1-\lambda_W)\lambda_\gamma x\right]\, ,
\end{equation}
\begin{equation}
{\cal M}^{CESM}_{\lambda_\gamma \, \, 0}-{\cal M}^{SM}_{\lambda_\gamma \, \, 0}=\frac{i\pi \alpha}{s_W}\, \alpha_W \, \left(\frac{s}{\Lambda^2}\right)\frac{\sqrt{x}\sqrt{1-x}\sin\left(\frac{\theta}{2}\right)}{1+x-(1-x)\cos\theta}\left\{(1+\lambda_\gamma)(1-x)+\left[
3-x-(1+x)\lambda_\gamma\right]\cos\theta\right\}\, .
\end{equation}
Notice that ${\cal M}^{CESM}_{+\, \, +}={\cal M}^{SM}_{+\, \, +}$, but  ${\cal M}^{CESM}_{\pm\, \, 0}\neq {\cal M}^{SM}_{\pm\, \, 0}$.

\subsection{New physics effects in the context of the SME}
\begin{figure}
\centering
\includegraphics[width=4.0in]{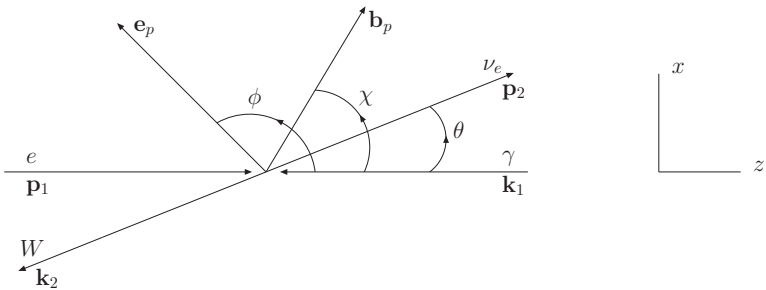}
\caption{\label{KandBF} The $\gamma e\to W\nu_e$ reaction in the c.m. frame in presence of the constant $b^{\alpha \beta}$ background tensor field. In the figure, $e_p$ and $b_p$ are the components of the $\textbf{e}$ and $\textbf{b}$ vectors on the collision plane.}
\end{figure}

We now turn to calculate the helicity amplitudes for the $e\gamma \to W\nu_e$ process in the presence of the constant background tensor $b^{\alpha \beta}$. The geometrical features of the collision are shown in Fig. \ref{KandBF}. In this figure, the electric-like, $e^i\equiv \Lambda^2_{LV} b^{0i}$, and and magnetic-like, $b^i\equiv (1/2)\Lambda^2_{LV}\epsilon^{ijk}b^{jk}$, constant fields have been decomposed into components parallel, $e_p$ and $b_p$, and perpendicular, $e_y$ and $b_y$, to the collision plane (the $x-z$ plane). The collision angle in the center of mass frame is denoted by $\theta$, whereas $\phi$ and $\chi$ are the angles formed by $e_p$ and $b_p$ with the $+z$ axes, respectively. We will need the following identity
\begin{eqnarray}
a_{\alpha}\, b^{\alpha\beta}
\,c_{\beta}&=&\frac{1}{\Lambda_{LV}^{2}}\,[c_{0}\,\textbf{e}\cdot \textbf{a}-a_{0}\,\textbf{e}\cdot
\textbf{c}+\textbf{b}\cdot
(\textbf{a}\times \textbf{c})]\, ,
\end{eqnarray}
valid for two arbitrary four-vectors $a_\mu$ and $c_\mu$.

The amplitude can be written as follows:

\begin{equation}
\mathcal{M}^{SME}_{\lambda_\gamma\, \, \bar{\lambda}_W}=\mathcal{M}^{SM}_{\lambda_\gamma\, \, \bar{\lambda}_W}+\frac{ie g}{2 \sqrt{2}} \,\left[ \overline u (p_2) \,
P_R\,\Gamma^{SME}_{\mu \nu} \, u(p_1)\right]\, \,
\varepsilon^{\mu}(k_1,\lambda_{\gamma})\,\,\varepsilon^{\ast
\nu}(k_2,\lambda_{\bar{W}})\, ,
\end{equation}
where
\begin{align}
\Gamma^{SME}_{\mu\nu}=&\frac{b^{\alpha\beta}\,}{2 \,(t-m_W^{2})} \Big \{ (m_W^{2}-t) (\gamma_{\mu } g_{\alpha \nu } k_{2\beta }-\gamma_{\mu } g_{\alpha \nu } k_{1 \beta }-\gamma_{\beta } k_{2\alpha } g_{\mu \nu })+(m_W^{2}+t)(\gamma_{\alpha } k_{1\beta } g_{\mu \nu }-\gamma_{\mu } k_{1\alpha } g_{\beta \nu })
   \nonumber\\
   +& \gamma_{\nu }
   \big [\left(m_W^{2}-t\right) \left(\left(k_{1\alpha
   }-k_{2\alpha }\right) g_{\beta \mu }+g_{\alpha \mu }
   k_{2\beta }\right)+4 k_{2\alpha } k_{1\beta }
   k_{2\mu }\big ]\nonumber\\
   +& 2 \big[k_{2\alpha } \left(2 \gamma_{\mu } k_{1\beta }+\gamma_{\beta } k_{2\mu
   }\right)+\gamma_{\alpha } \left(k_{1\beta } k_{2\mu
   }-m_W^{2} g_{\beta \mu }\right)\big ] k_{1\nu }\nonumber\\
   +& 2
   \pFMSlash{k}_1 \big [-2 k_{2\alpha }
   k_{1\beta } g_{\mu \nu }+\left(k_{1\alpha }
   g_{\beta \nu }+g_{\alpha \nu } \left(k_{1\beta
   }-k_{2\beta }\right)\right) k_{2\mu}-\left(k_{1\alpha }-k_{2\alpha }\right) g_{\beta
   \mu } k_{1\nu }+g_{\alpha \mu } \left(t g_{\beta \nu
   }+k_{2\beta } k_{1\nu }\right)\big ] \Big \}
\end{align}

After a tedious algebra, the corresponding helicity amplitudes can be written as follows
\begin{equation}
\mathcal{M}^{SME}_
{\tiny{\lambda \gamma\,\lambda_W}}-\mathcal{M}^{SM}_
{\tiny{\lambda \gamma\,\lambda_W}}= \left(\frac{\pi\alpha}{16\sqrt{2}s_W}\right)\left(\frac{s}{\Lambda^2_{LV}}\right)\frac{\sqrt{1-x}}{1+x-(1-x)\cos\theta}\left[ E^{\lambda_W}_y \, e_y+B^{\lambda_W}_p \,b_p
+i \, (E^{\lambda_W}_p\,e_p+B^{\lambda_W}_y \,b_y) \right],
\end{equation}
\begin{equation}
\mathcal{M}^{SME}_
{\tiny{\lambda \gamma\,\lambda_W^0}}-\mathcal{M}^{SM}_
{\tiny{\lambda \gamma\,\lambda_W^0}}=\left(\frac{\pi\alpha}{16s_W}\right)\left(\frac{s}{\Lambda^2_{LV}}\right)\frac{\sqrt{1-x}}
{\sqrt{x}\,(1+x-(1-x)\cos\theta)}\left[
E^{\lambda_W^0}_y \, e_y+B^{\lambda_W^0}_p \, b_p+i\,
(E^{\lambda_W^0}_p \, e_p+B^{\lambda_W^0}_y \, b_y) \right],
\end{equation}
where
\begin{eqnarray}
E^{\lambda_W}_y&=&4 \Big \{-\lambda_W+\lambda \gamma +x [\lambda_W x+x+4 \lambda_W+(2 x+4 (x+3) \lambda_W+7) \lambda \gamma -1]\nonumber \\
   &+&(\lambda_W-\lambda
   \gamma +x (\lambda_W x+x-2 \lambda_W+(2 x+4 (x-1) \lambda_W+3) \lambda \gamma
   +5)) \cos (\theta ) \Big\} \sin \left(\frac{\theta }{2}\right),
\end{eqnarray}
\begin{eqnarray}
B^{\lambda_W}_p&=&-2  \Big\{\big[2 (4
   \lambda \gamma  \lambda_W+\lambda_W+2
   \lambda \gamma +1) x^2+(4 \lambda \gamma  \lambda_W+2 \lambda_W+9 \lambda \gamma -11) x-5
   \lambda \gamma \nonumber \\
   &-&4 (3 \lambda \gamma \lambda_W+\lambda_W)+3\big] \sin (\theta -\chi
   )+(x-1) [\lambda_W x+x-\lambda_W+2
   (x-1) (2 \lambda_W+1) \lambda \gamma +1] \sin
   (2 \theta -\chi )\nonumber \\
   &-&[(x-1) (x+(x-5) \lambda_W+3)+2 (6 \lambda_W+x (x+2 (x-4)
   \lambda_W+3)+2) \lambda \gamma ] \sin (\chi
   )\nonumber \\
   &-&[2 \lambda_W x+x-2 \lambda_W+(x-1)
   (4 \lambda_W+1) \lambda \gamma +1] \sin (\theta
   +\chi )\Big\}\sin \left(\frac{\theta }{2}\right),
\end{eqnarray}
\begin{eqnarray}
E^{\lambda_W}_p&=&-2  \Big[2 (x+3) ((x-1)
   \lambda_W-2)-(-3 \lambda_W+x (x+(x+18)
   \lambda_W+12)+3) \lambda \gamma \nonumber \\
   &+&4 (-\lambda
   \gamma  \lambda_W+2 \lambda_W-2 x
   (\lambda_W+2)+x (\lambda_W+3) \lambda
   \gamma +\lambda \gamma +4) \cos (\theta )+(x-1) ((x-1)
   \lambda_W (\lambda \gamma -2)\nonumber \\
   &+&x \lambda \gamma
   +\lambda \gamma +4) \cos (2 \theta )\Big] \cos (\phi )\cos \left(\frac{\theta }{2}\right)-2 \Big[-2
   x ((x+5) \lambda_W+12)+(-\lambda_W+x
   (x+(x-12) \lambda_W-4)+3) \lambda \gamma \nonumber \\
   &+&2
   (\lambda_W-\lambda \gamma +x (-2 x
   \lambda_W+\lambda_W+(x+(x-1)
   \lambda_W+4) \lambda \gamma -6)) \cos (\theta
   )\nonumber \\
   &+&(x-1) ((x-1) \lambda_W (\lambda \gamma -2)+x
   \lambda \gamma +\lambda \gamma +4) \cos (2 \theta )+4\Big]
   \sin \left(\frac{\theta }{2}\right) \sin (\phi )
\end{eqnarray}
\begin{eqnarray}
B^{\lambda_W}_y&=&4 \Big\{[\lambda_W (\lambda \gamma -2)+\lambda
   \gamma ] x^2+[2 (\lambda \gamma -1)+3 \lambda_W
   (\lambda \gamma +1)] x\nonumber \\
   &-&3 \lambda_W-2
   \lambda_W \lambda \gamma -3 \lambda \gamma \big[(\lambda_W (\lambda \gamma -2)+\lambda
   \gamma ) x^2-(3 \lambda \gamma  \lambda_W+\lambda_W+4 \lambda \gamma +6) x\nonumber \\
   &+&3
   \lambda_W+2 \lambda_W \lambda \gamma
   +3 \lambda \gamma +4 \big] \cos (\theta )-4 \Big\} \sin
   \left(\frac{\theta }{2}\right),
\end{eqnarray}
\begin{equation}
E^{\lambda_W^0}_y= 4x  (-\lambda \gamma +x
   ((2 x+5) \lambda \gamma -4)+(\lambda \gamma +x (2 x
   \lambda \gamma +\lambda \gamma +4)+2) \cos (\theta )-2)\cos \left(\frac{\theta }{2}\right),
\end{equation}
\begin{align}
B^{\lambda_W^0}_p=& 4x \Big\{[x (x \lambda
   \gamma +\lambda \gamma +3)+(-\lambda \gamma +x (2 x
   \lambda \gamma +3 \lambda \gamma -4)+2) \cos (\theta
   )\nonumber \\
   +&(x-1) ((x-1) \lambda \gamma +1) \cos (2 \theta )-1]
   \sin (\chi ) \cos \left(\frac{\theta }{2}\right) - [2 (\lambda \gamma +x (2 x \lambda \gamma
   -\lambda \gamma -1)-1) \cos (\theta )\nonumber \\
   +&(x-1) (3 (x
   \lambda \gamma +\lambda \gamma -1)+((x-1) \lambda \gamma
   +1) \cos (2 \theta ))] \cos (\chi ) \sin
   \left(\frac{\theta }{2}\right)\Big\},
\end{align}
\begin{eqnarray}
E^{\lambda_W^0}_p&=&4 x  \big[\lambda \gamma +x
   (x+3 \lambda \gamma +7)+(x+1) (2 x-2 \lambda \gamma +1)
   \cos (\theta )\nonumber \\
   &+&(x-1) (x-\lambda \gamma -3) \cos (2
   \theta )-4 \big] \sin (\phi )\cos \left(\frac{\theta }{2}\right)\nonumber \\
   &-&4 x \big[3 x (x-\lambda \gamma )+3
   \lambda \gamma +4 ((x-1) x-(x+1) \lambda \gamma ) \cos
   (\theta )\nonumber \\
   &+&(x-1) (x-\lambda \gamma -3) \cos (2 \theta
   )+5\big] \cos (\phi ) \sin \left(\frac{\theta }{2}\right),
\end{eqnarray}
\begin{align}
B^{\lambda_W^0}_y=&-4 x \left(2 x^2-2
   \lambda \gamma  x+x+2 \lambda \gamma +\left(2 x^2+x+2
   (x-1) \lambda \gamma -1\right) \cos (\theta )+1\right) \cos \left(\frac{\theta }{2}\right).
\end{align}

As already commented, we are interested not only in investigating possible deviations on the SM prediction for the $e\gamma \to W\nu_e$ reaction in both the SME and the CESM contexts, but also in comparing among themselves these sources of new physics.

\section{Discussion}\label{D}
In this section, we analyze our results. It can be seen from Eqs.~(\ref{SMAMP1})-(\ref{SMAMP2}) that there exist six polarization states, namely: $(-,-)$, $(-,+)$, $(-,0)$, $(+,-)$, $(+,+)$, and $(+,0)$. Although the $(-,+)$ and $(-,0)$ SM polarized amplitudes are exactly zero at the tree-level, these polarized amplitudes receive contributions at the one-loop level~\cite{RSM}. However, we will not consider these contributions by simplicity, concentrating our attention in only tree-level effects.

The new physics that does not respect the Lorentz symmetry gives additional information which could be more interesting than the new physics respecting this symmetry. The former can be evidenced not only by the relative value of the new physics scale but also due to privileged directions determined by the $\textbf{b}, \textbf{e}$ constant background fields. In order to search for possible scenarios of violation of Lorentz symmetry we will study in detail the differential cross section for the $e \gamma \to W\nu_e$ process. It is important to stress that simultaneous information about dependency of privileged angular directions of the background fields and scattering angle in the total cross section is lost, since the scattering angle has been already integrated. Hence, we present an exhaustive study of the differential cross section. One of the questions we want to answer is whether Lorentz violation is more sensitive to the $\textbf{b}$ or to the $\textbf{e}$ background fields. We are interested in the search for scenarios in which either the SM contribution is absent or the new physics effects differ significantly from it. In this sense, we look for optimal values for the Lorentz violating parameters which enhance the new physics effects arising from the SME. To this end, we will analyze three different scenarios, namely: a) $\textbf{e}=0$, $\textbf{b}\neq 0$, b) $\textbf{e}\neq0$, $\textbf{b}=0$, and c) $\textbf{e}\neq0$, $\textbf{b}\neq 0$.

As it has already emphasized, we are interested in contrasting new physics effects arising in the CESM approach or in SME one, since they could be observed at ILC. To make predictions, some values for the parameters of the CESM, $(\Lambda, \alpha_W)$, and for the ones of the SME, $(\Lambda_{LV}, e_p, e_y, b_p, b_y, \chi, \phi)$, must be assumed. In a previous work by some of us, a constraint given by $\Lambda_{LV}>1.96$ TeV on the Lorentz-violating scale associated with the ${\cal O}^W$ operator, has been derived from experimental data on the $B\to X_s\gamma$ decay~\cite{T1}. Due to this and also for comparison purposes of both types of effective theories, in the following we will assume that $\Lambda=\Lambda_{LV}=2$ TeV. In addition, we will assume that $\alpha_W=1$.

\subsection{Differential cross section}
Because the experimental restrictions, we will discuss our analysis of the $e \gamma \to W\nu_e$ differential cross section on the scattering angle interval $20^\circ<\theta<160^\circ$, which is consistent with the kinematical cuts introduced in Ref.~\cite{RSM}. We will only analyze the behavior of the $(+,+), (+,-), (-,+), (-,0)$ polarized differential cross sections, as the new physics effects on the $(-,-), (+,0)$ ones are marginal.

\subsubsection{Scenario $\textbf{e}=0$, $\textbf{b}\neq 0$}
In all cases, we will use the values $b_p=b_y=1$ for the $\textbf{b}$ constant field.

\begin{figure}[htb!]
\centering
\subfigure[]{\includegraphics[scale=0.65]{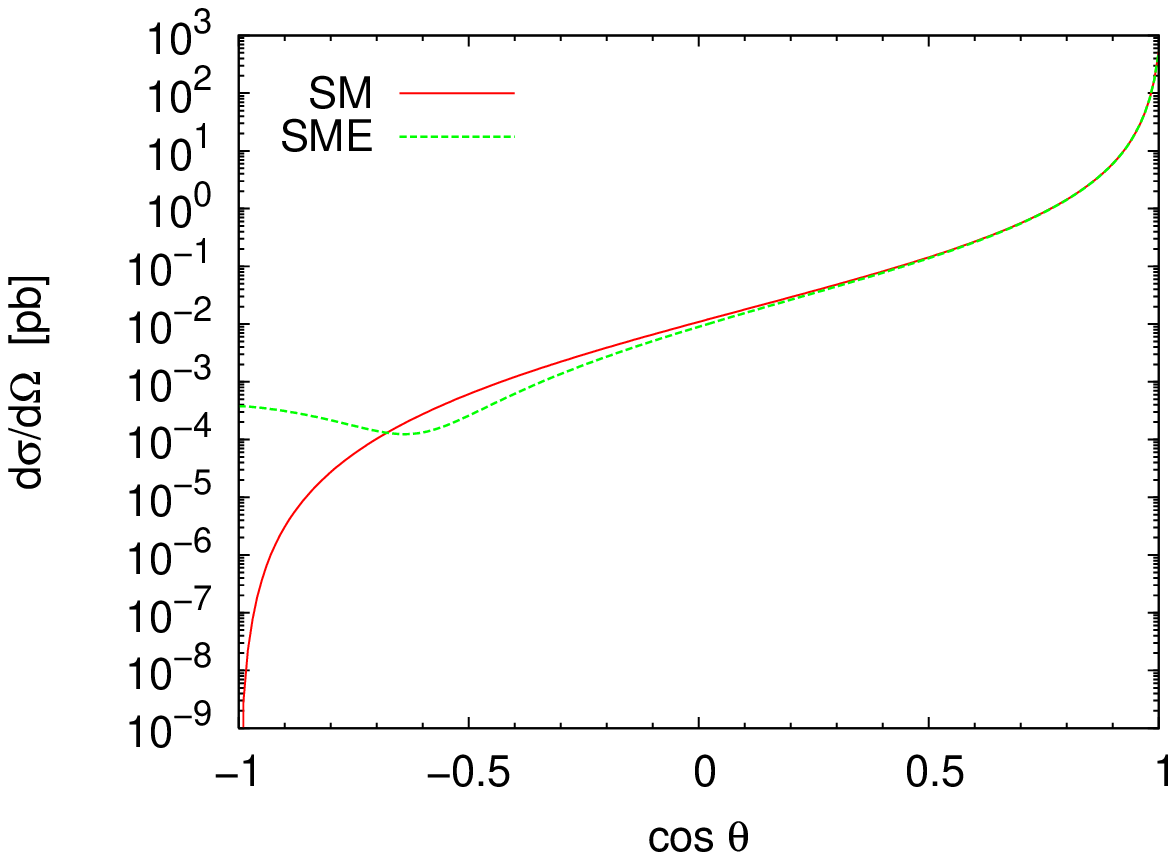}}\qquad
\subfigure[]{\includegraphics[scale=0.65]{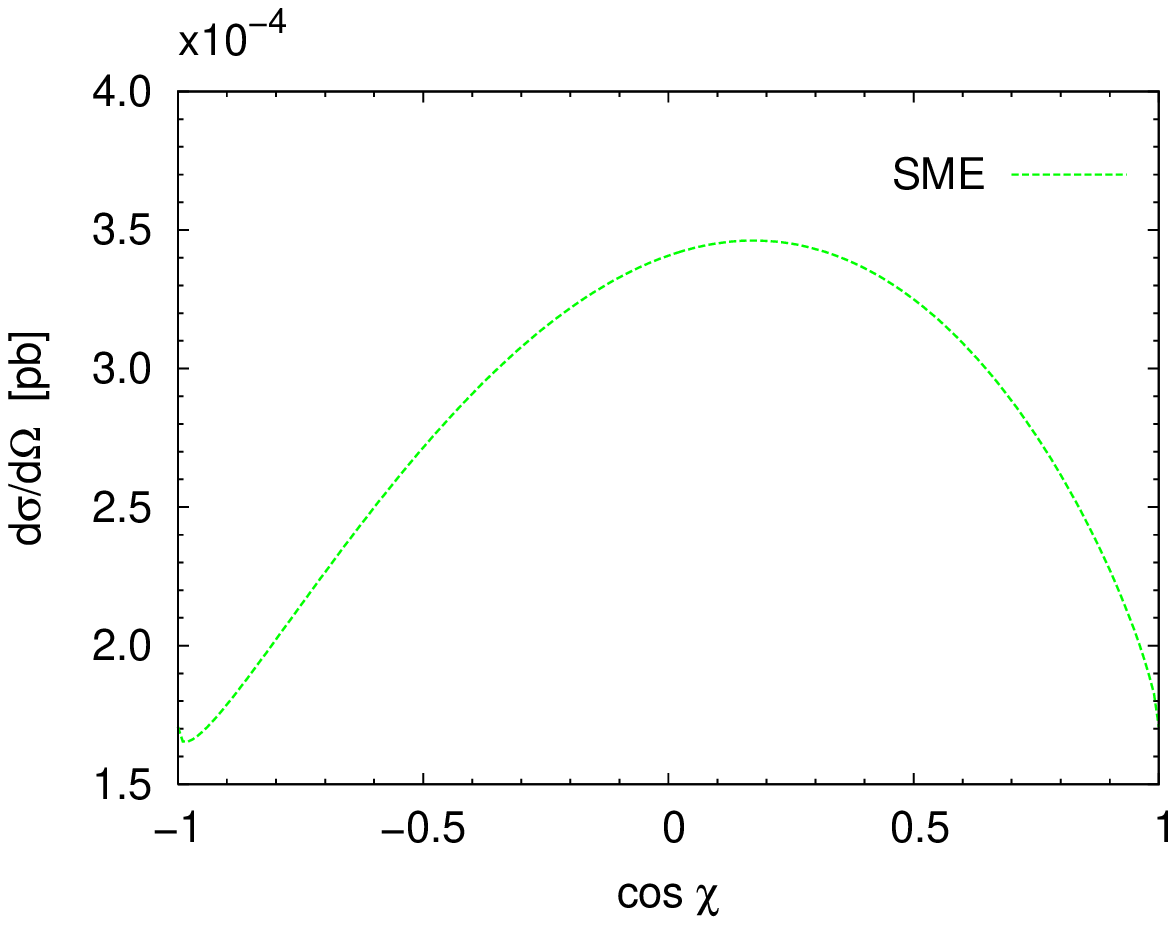}}
\caption{Differential cross section for the $e \gamma \to W\nu_e$ process with the $(+,+)$ polarization state at $\sqrt{s}=1$ TeV ($\textbf{e}=0$, $\textbf{b}\neq 0$). (a) $\chi=78.46^\circ$. (b) $\theta=160^\circ$.}
\label{dcsppang}
\end{figure}

\noindent \textbf{The $(+,+)$ collision.} In this type of collision there is no anomalous contribution from CESM, so the new physics effects arise exclusively from the SME. In consequence, this is a good scenario to detect signals of Lorentz violation. The polarized differential cross section as a function of the scattering angle and the $b_p$ angular direction $\chi$ is presented in Fig.~\ref{dcsppang}. In Fig.~\ref{dcsppang}(a), the differential cross sections for both the SM and the SME predictions as a function of $\cos\theta$ are displayed. It can be appreciated from this figure that the Lorentz violating effect is about two orders of magnitude larger than the SM prediction for $\chi=78.46^\circ$ and $\theta= 160^\circ$. Fig.~\ref{dcsppang}(b) shows the differential cross section as a function of $\cos\chi$. It can be appreciated from this figure that the differential cross section reach its maximum value for  $\chi=78.46^\circ$. From both figures one can to conclude that Lorentz violation becomes more intense in the neighborhood of $\theta=160^\circ$ and $\chi=78.46^\circ$. \textit{The main feature of this collision is that only physics whose source is Lorentz violation can show up.}

\begin{figure}[htb!]
\centering
\subfigure[]{\includegraphics[scale=0.65]{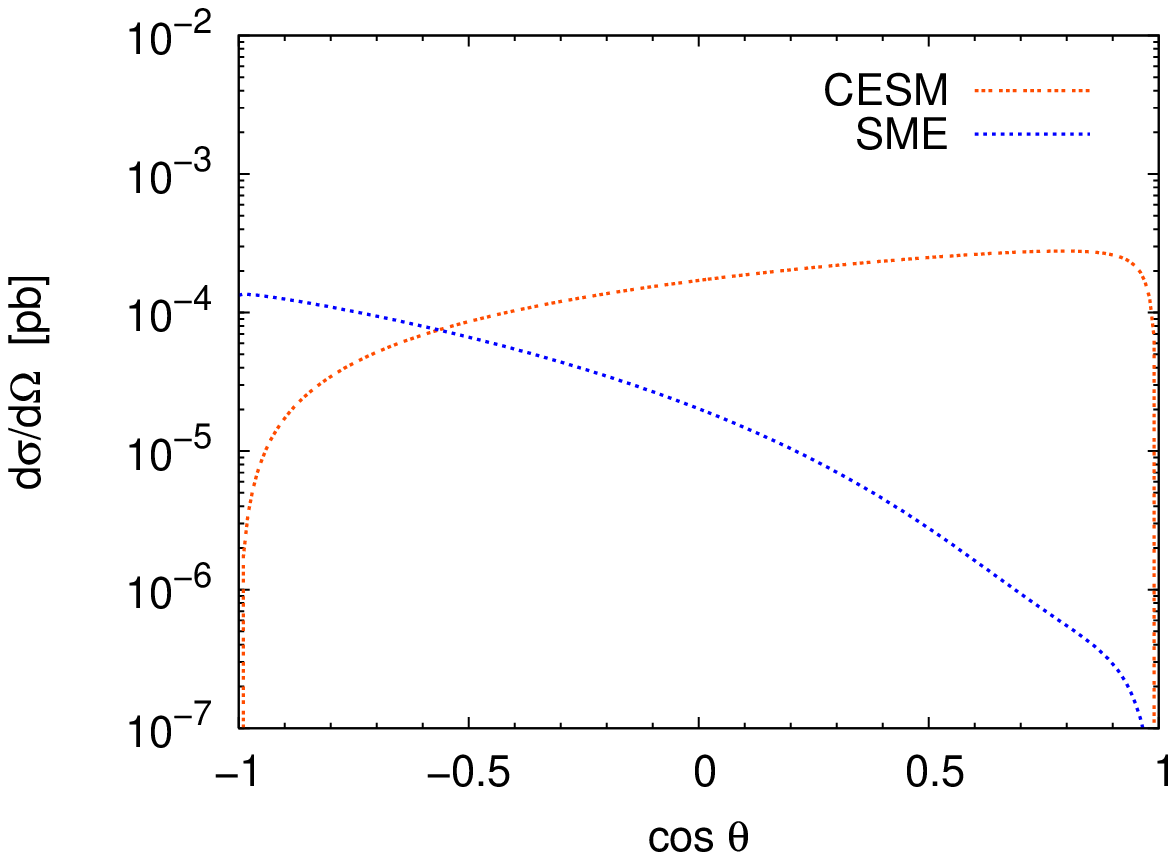}}\qquad
\subfigure[]{\includegraphics[scale=0.65]{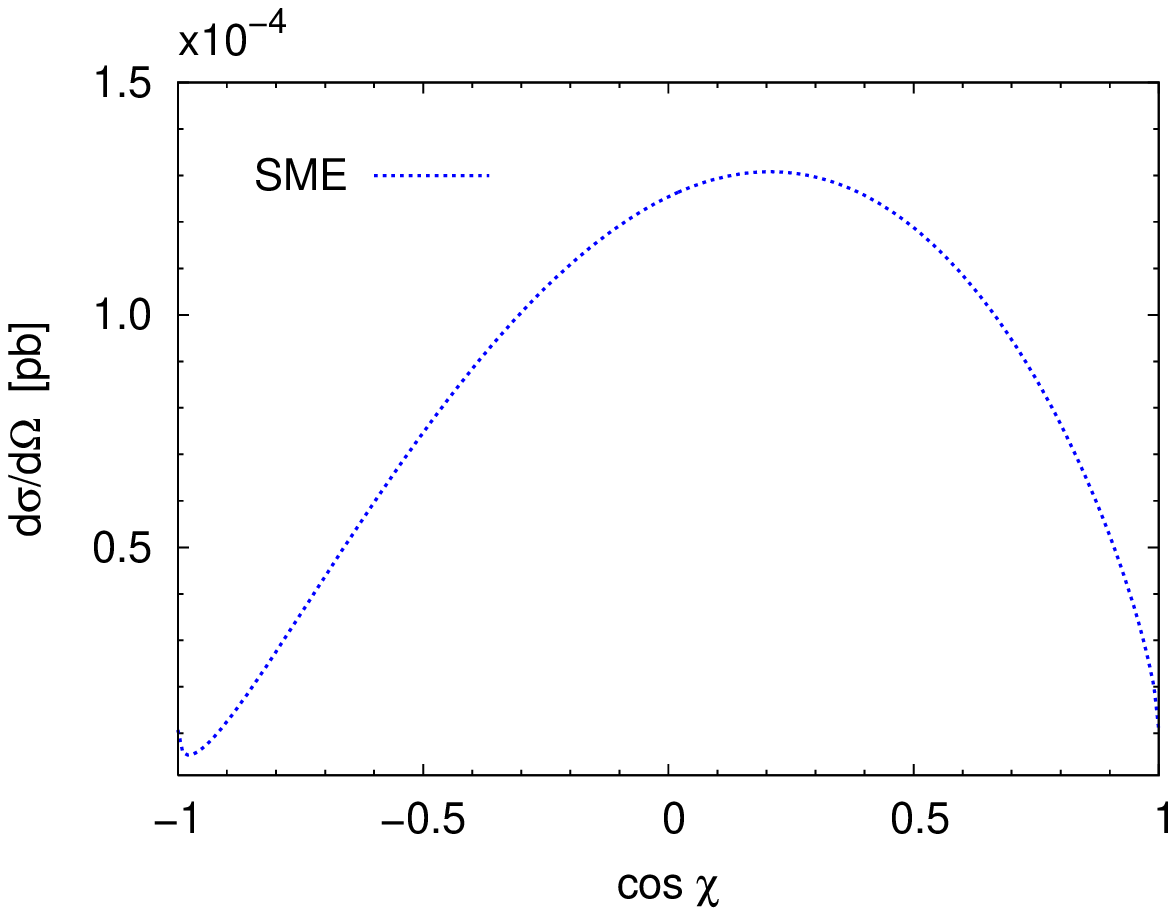}}
\caption{Differential cross section for the $e \gamma \to W\nu_e$ process with the $(-,+)$ polarization state at $\sqrt{s}=1$ TeV ($\textbf{e}=0$, $\textbf{b}\neq 0$), where there is no contribution of the SM. (a) $\chi=78.46^\circ$. (b) $\theta=160^\circ$.}
\label{dcsmpang}
\end{figure}

\noindent \textbf{The $(-,+)$ collision.} The main feature of this collision is that the SM contribution vanishes exactly at this order of perturbation theory. Therefore, it is a good scenario to confront new physics effects arising from CESM and SME. In Fig.~\ref{dcsmpang}, the $(-,+)$ polarized differential cross section is presented as a function of the scattering angle and the $b_p$ angular direction $\chi$. It can be appreciated from this figure a quite different behavior of the two types of new physics effects. In Fig.~\ref{dcsmpang}(a), which displays the differential cross section as a function of $\cos\theta$, it can be observed that the CESM contribution is larger than the SME one by about $3$ orders of magnitude for $\theta= 25^\circ$. Nevertheless, when $\theta=155^\circ$, the SME contribution is larger than the CESM contribution by approximately 1 order of magnitude. In Fig.~\ref{dcsmpang}(b), it can be appreciated that new physics effects arising from SME are favored for angular directions in the neighborhood of  $\chi=78.46^\circ$ and $\theta=160^\circ$. \textit{Detection of new physics effects is favored by this collision, as the SM contribution first arise at the one-loop level. Signals coming from SME and CESM are clearly distinguished for some angular regions.}

\begin{figure}[htb!]
\centering
\subfigure[]{\includegraphics[scale=0.65]{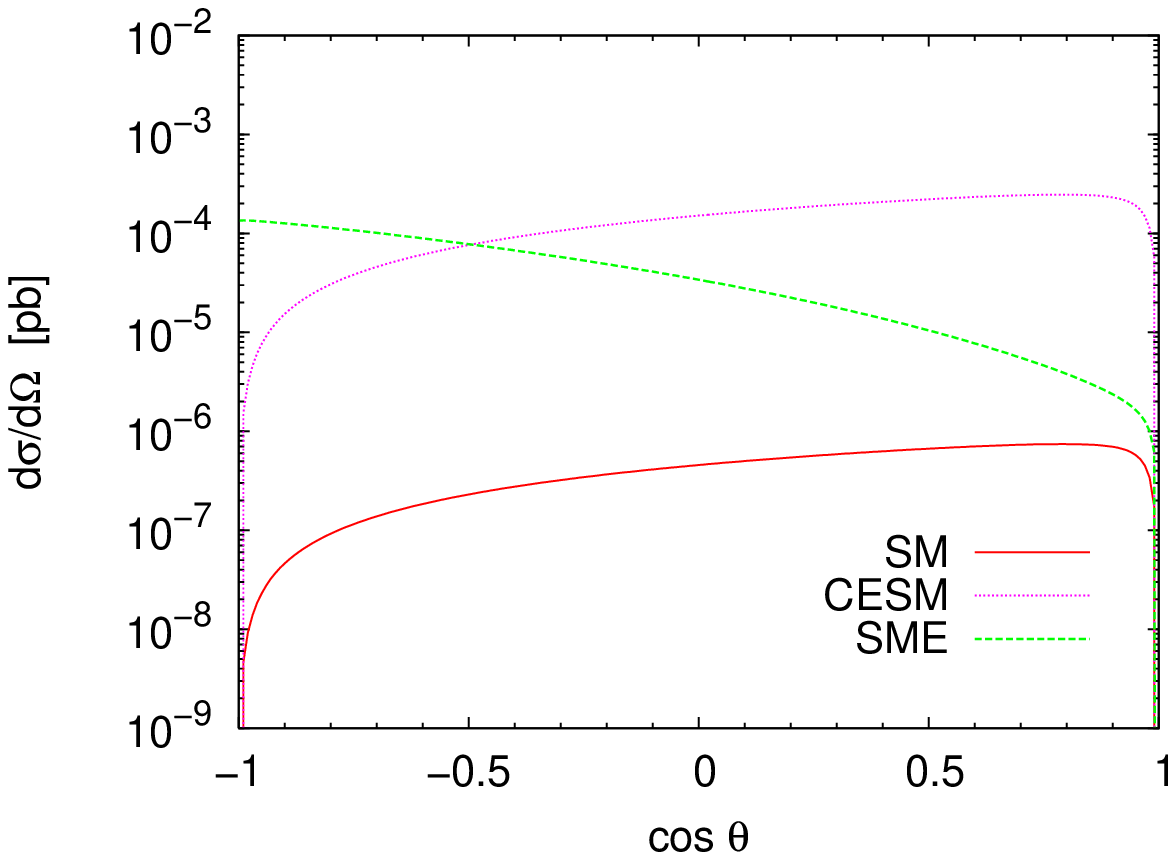}}\qquad
\subfigure[]{\includegraphics[scale=0.65]{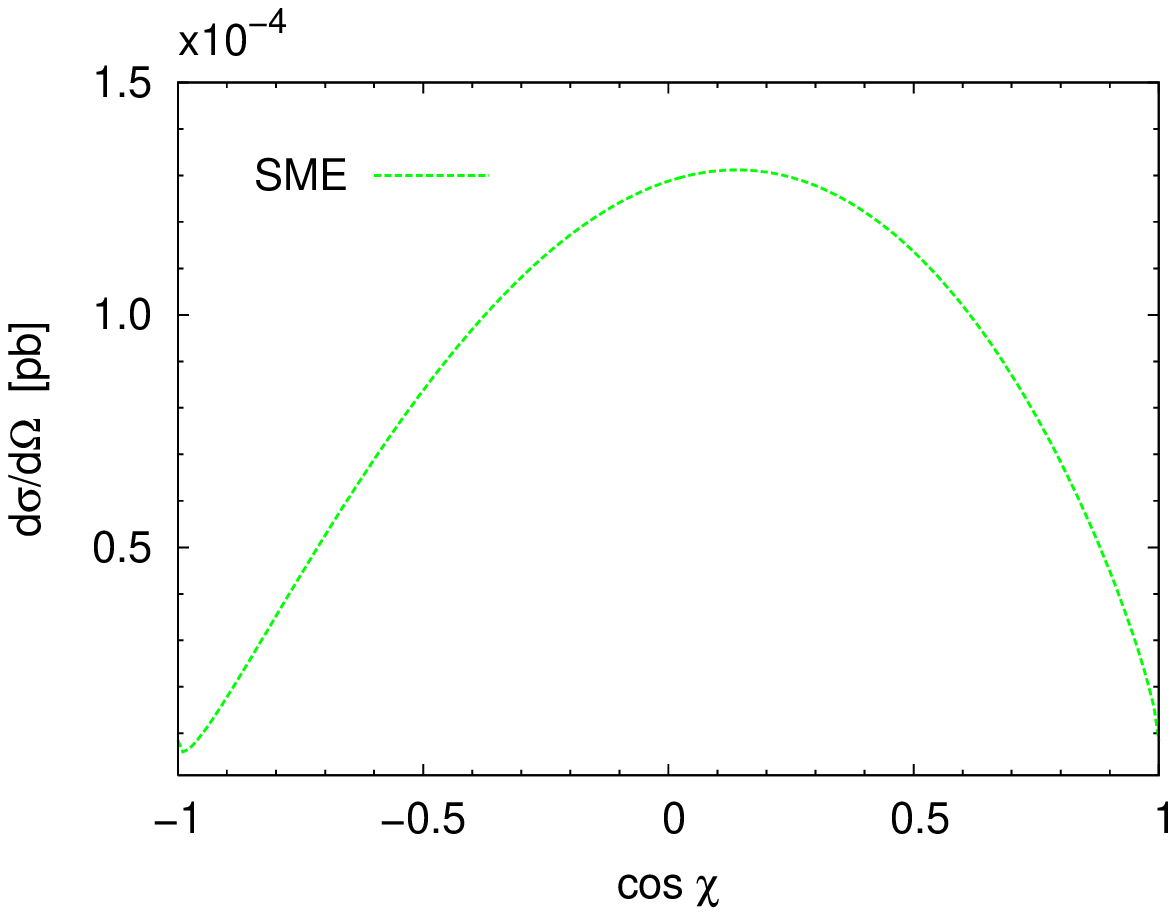}}
\caption{Differential cross section for the $e \gamma \to W\nu_e$ process with the $(+,-)$ polarization state at $\sqrt{s}=1$ TeV ($\textbf{e}=0$, $\textbf{b}\neq 0$). (a) $\chi=79.63^\circ$. (b) $\theta=160^\circ$.}
\label{dcspmang}
\end{figure}

\noindent \textbf{The  $(+,-)$ collision.} In Fig.~\ref{dcspmang} we display the $(+,-)$ polarized differential cross section as a function of the scattering angle and the $b_p$ angular direction $\chi$. In this case, the SM contribution is subdominant. It can be appreciated from this figure a quite different behavior between the SME and CESM contributions. In Fig.~\ref{dcspmang}(a), the differential cross section as a function of $\cos\theta$ is displayed, showing that the gap between SME and SM contribution is maximized as the scattering angle tends to $180^\circ$. In particular, for $\chi=78.46^\circ$ and $\theta= 155^\circ$ the SME contribution dominates in around 3 orders of magnitude. For the same scattering angle, the CESM contribution also is larger than the SM one in around 2 orders of magnitude. However, the CESM contribution is larger than the SME one about $2$ orders of magnitude for $\theta= 25^\circ$. Once again, in Fig.~\ref{dcspmang}(b) we can see that the preferential angular direction for physics coming from SME occurs in the neighborhood of $\chi=78.46^\circ$ for $\theta=160^\circ$. \textit{The peculiarity of this collision is that the SM contributions is subdominant with respect to new physics effects arising from both the SME and CESM and that both sources of physics beyond the SM can be distinguished in some angular regions.}

\begin{figure}[htb!]
\centering
\subfigure[]{\includegraphics[scale=0.65]{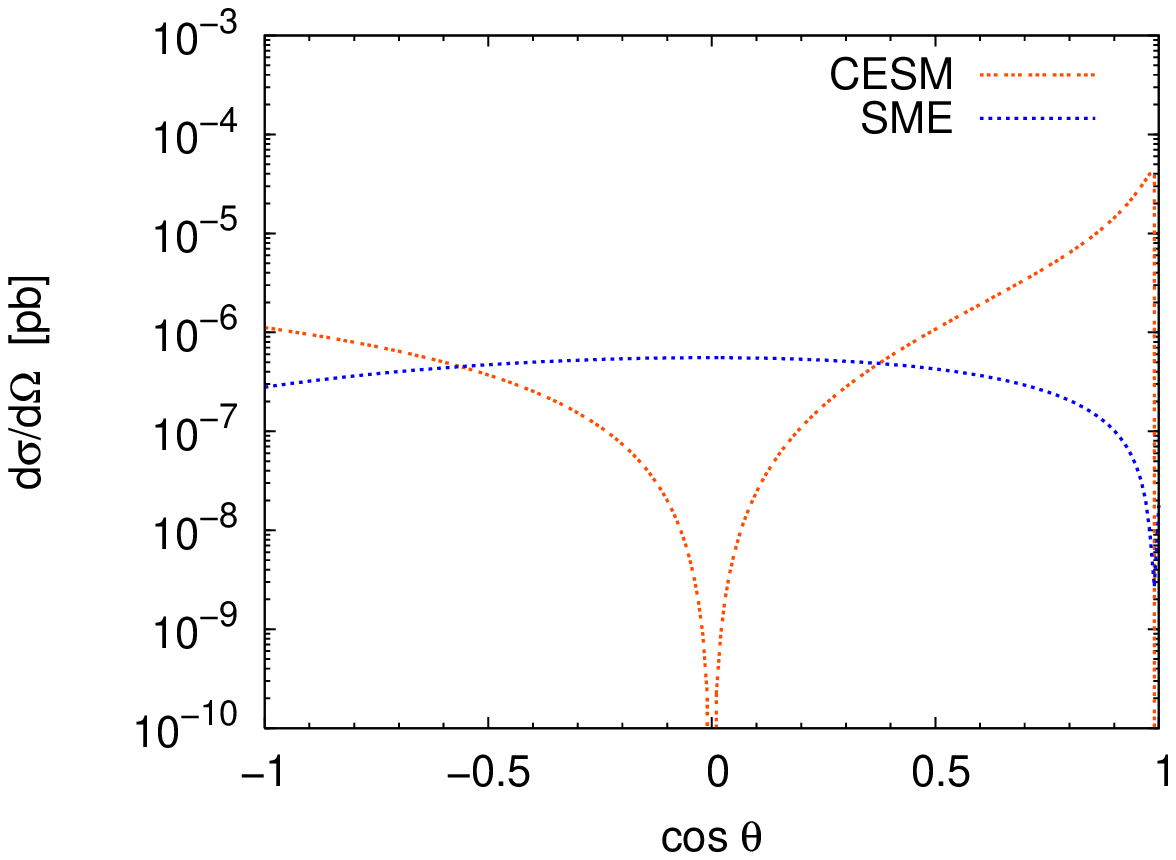}}\qquad
\subfigure[]{\includegraphics[scale=0.65]{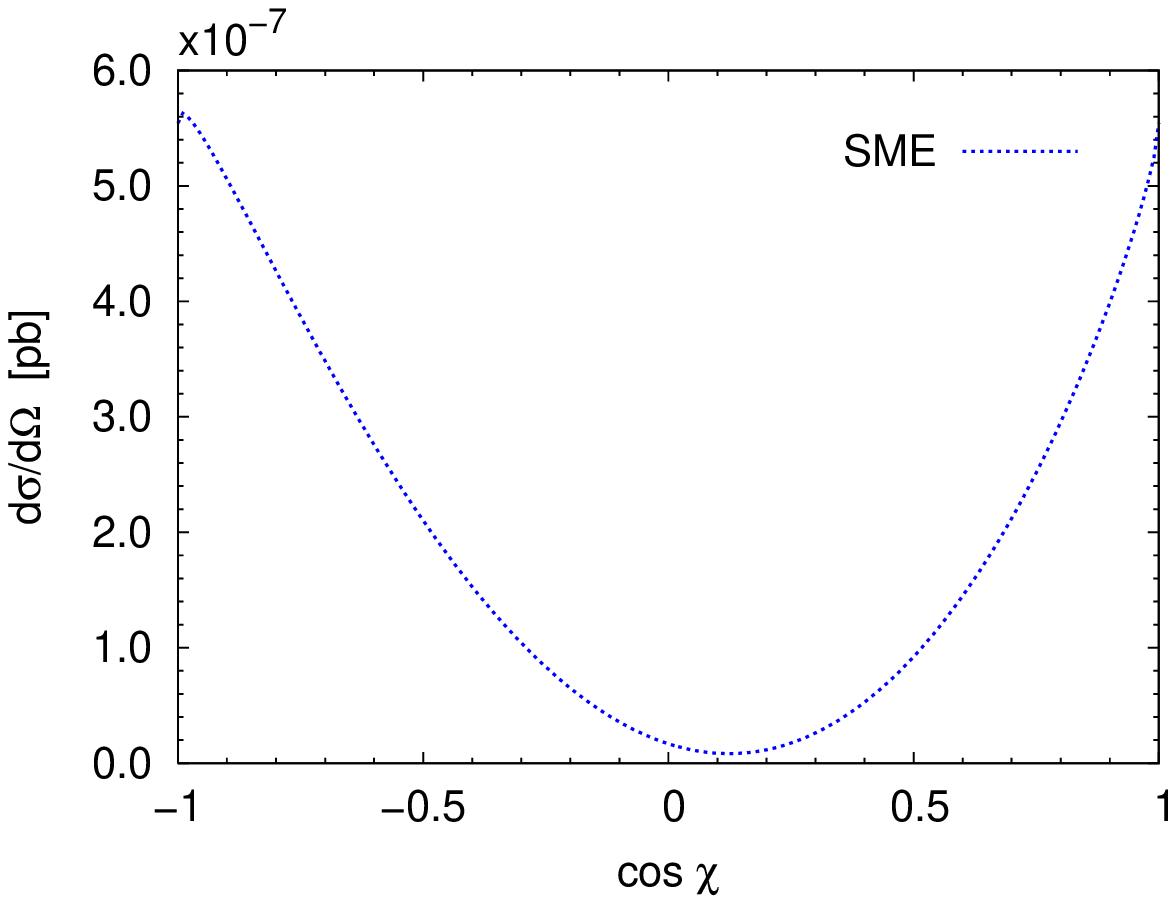}}
\caption{Differential cross section for the $e \gamma \to W\nu_e$ process with the $(-,0)$ polarization state at $\sqrt{s}=1$ TeV ($\textbf{e}=0$, $\textbf{b}\neq 0$), where there is no contribution of the SM. (a) $\chi=0^\circ$. (b) $\theta=90^\circ$.}
\label{dcsmang}
\end{figure}

\noindent \textbf{The $(-,0)$ collision.} This collision is a good mode to search new physics effects, as the SM contributions vanishes at the tree level. In Fig.~\ref{dcsmang} the $(-,0)$ polarized differential cross section as a function of the scattering angle and the $b_p$ angular direction $\chi$ is shown. It can be appreciated from this figure a completely different behavior between the SME and CESM contributions. In particular, the CESM contribution is exactly zero for $\theta=90^\circ$. From Fig.~\ref{dcsmang}(a), which displays the differential cross section as a function of $\cos\theta$, it can be appreciated that Lorentz violation could be clearly observed just in the neighborhood of $\theta=90^\circ$. Fig.~\ref{dcsmang}(b) shows that an optimal detection of Lorentz violation signal could occurs in the the neighborhood of $\chi= 25^\circ$ or $\chi= 155^\circ$.

To finish, we would like to summarize our results concerning the impact of Lorentz violation characterized by the presence of a constant magnetic-like  background field. We focus on the optimum scenario to search for signals of Lorentz violation. Lorentz violation effects show up clearly in the $(+,+)$ collision just in the neighborhood of  $\theta=160^\circ$ and  $\chi\simeq 79^\circ$. The corresponding differential cross section is about $10^{-4}$ pb, being the SM prediction about 2 orders of magnitude lower. As far as the $(-,+)$ collision is concerned, the Lorentz violation signal is also clearest in the neighborhood of  $\theta=160^\circ$ and  $\chi\simeq 79^\circ$. The differential cross section is also of order of $10^{-4}$ pb. The main feature of this case is that the SM contribution is exactly zero at this order of perturbation theory, magnifying thus signals of new physics. On the other hand, the $(+,-)$ collision shows as, in the two previous cases, signals of Lorentz violation are favored in the neighborhood of  $\theta=160^\circ$ and  $\chi\simeq 79^\circ$, also with a differential cross section of order of $10^{-4}$ pb. In this case, the SM contribution is subdominant for all scattering angles. Concerning the $(-,0)$ collision, it shows an atypically behavior, as a Lorentz violation signal is favored in the neighborhood of  $\theta=90^\circ$ and  $\chi\simeq 0^\circ$. Although the differential cross section is lower than the other collisions by about two orders of magnitude, it is important to notice that it is a very clear signal because the SM contributions vanishes exactly at this order of perturbation theory.
%%%%%%%%%%%%%%%%%%%%%%%%%%%%%%%%%%%%%%%%%%%%%%%%%%%%%%%%%%%%%%%%%%%%%%%%%%%%%%%%%%%%%%%

\begin{figure}[htb!]
\centering
\subfigure[]{\includegraphics[scale=0.65]{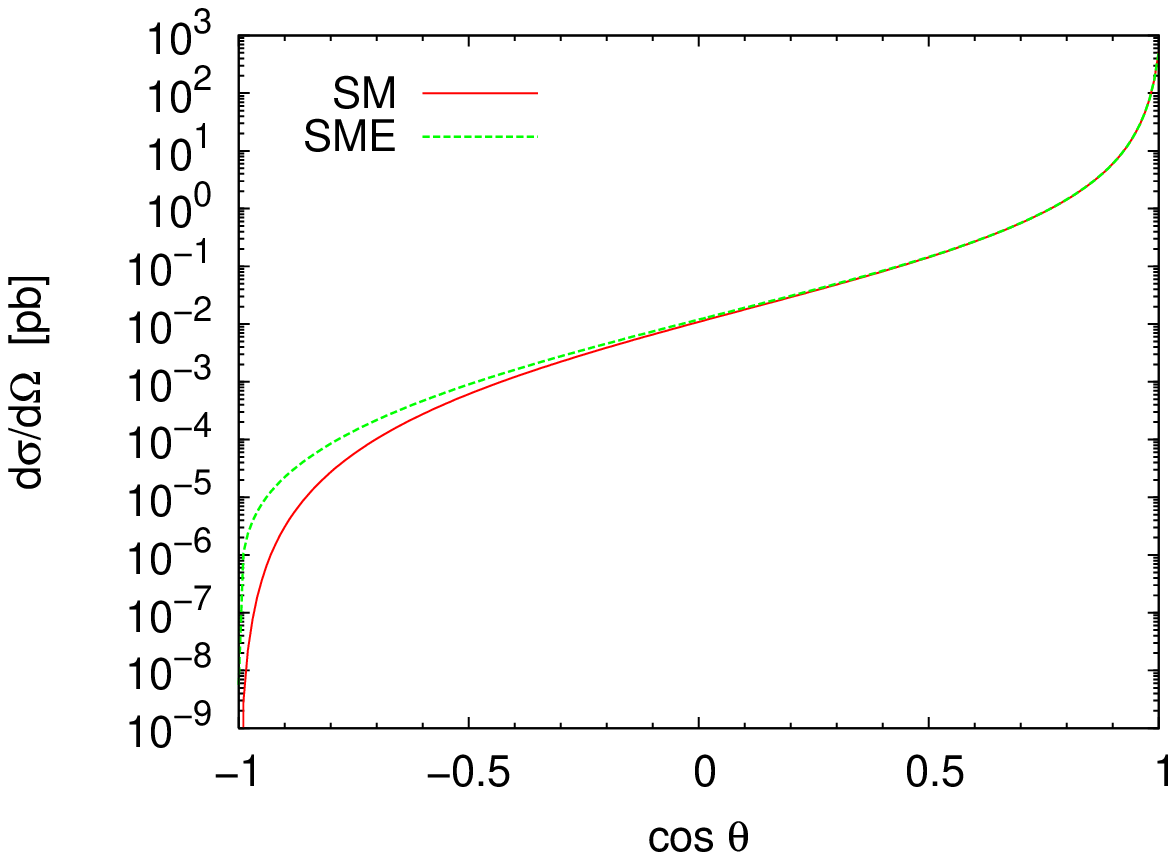}}\qquad
\subfigure[]{\includegraphics[scale=0.65]{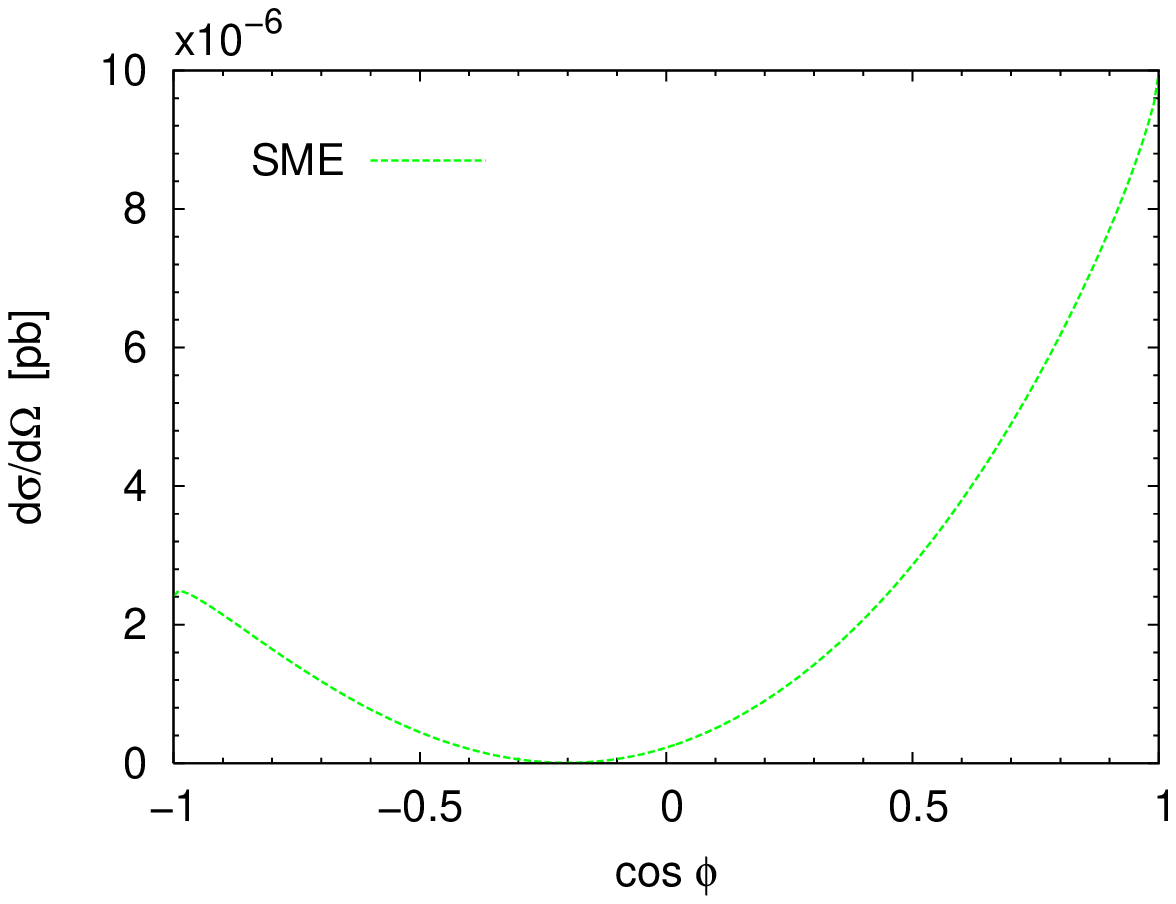}}
\caption{Differential cross section for the $e \gamma \to W\nu_e$ process with the $(+,+)$ polarization state at $\sqrt{s}=1$ TeV ($\textbf{e}\neq0$, $\textbf{b}=0$). (a) $\phi=0^\circ$. (b) $\theta=160^\circ$.}
\label{dcsppangb0}
\end{figure}

\subsubsection{Scenario $\textbf{e}\neq0$, $\textbf{b}=0$}
In this scenario we will use $e_p=e_y=1$ for the parallel and normal components of the constant background field $\textbf{e}$.

\noindent \textbf{The $(+,+)$ collision.} It should be recalled that there is no contributions from CESM to this collision. In Fig.~\ref{dcsppangb0}, the $(+,+)$ polarized differential cross section is shown as a function of the scattering angle and the $e_p$ angular direction $\phi$. From this figure, it can be appreciated that the differential cross section is of the order of  $10^{-6}$ pb in the neighborhood of $\chi=78.46^\circ$ for $\theta=160^\circ$, at the best. This in contrast with the scenario $\textbf{e}=0$, $\textbf{b}\neq 0$, in which the corresponding differential cross section is about of $10^{-4}$ pb. As it can be appreciated from theses figures, the Lorentz violating effect is not significant in this case. \textit{This collision is essentially insensitive to new physics effects.}

\begin{figure}[htb!]
\centering
\subfigure[]{\includegraphics[scale=0.65]{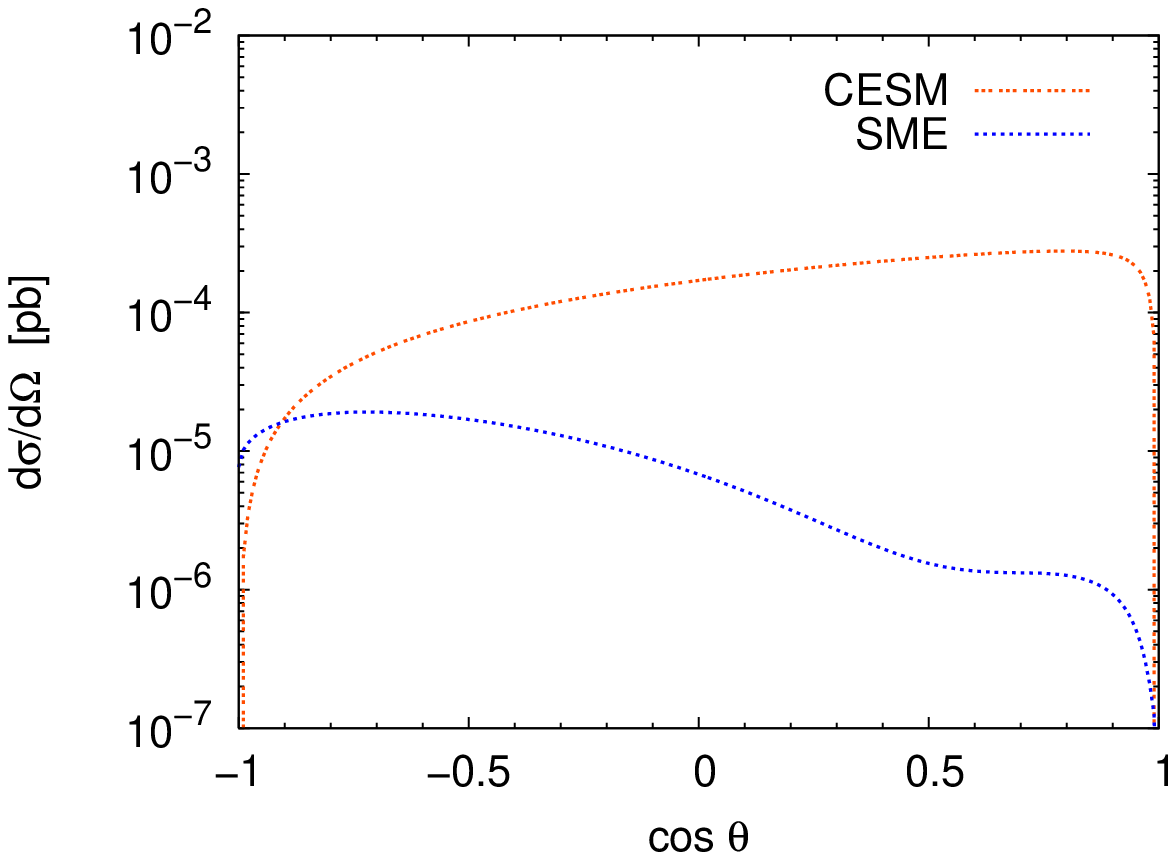}}\qquad
\subfigure[]{\includegraphics[scale=0.65]{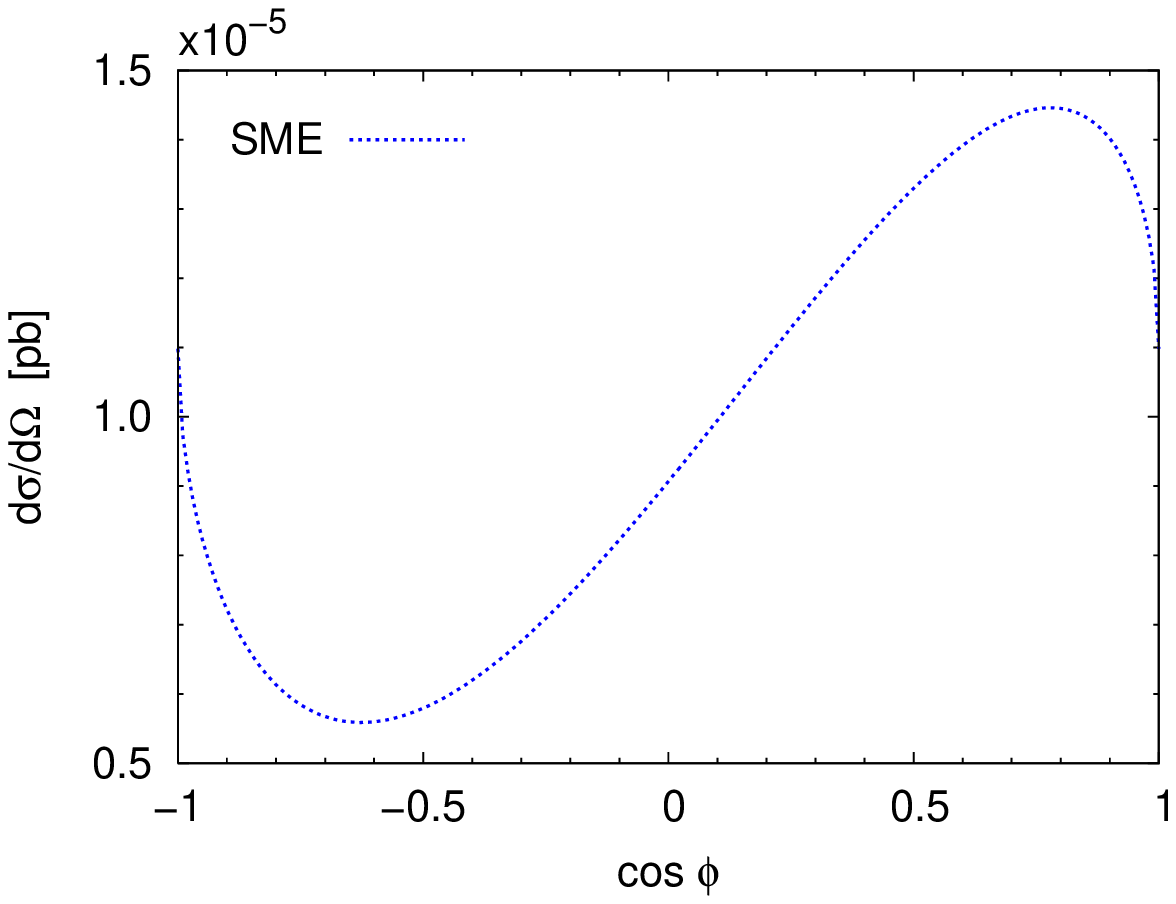}}
\caption{Differential cross section for the $e \gamma \to W\nu_e$ process with the $(-,+)$ polarization state at $\sqrt{s}=1$ TeV ($\textbf{e}\neq0$, $\textbf{b}= 0$), where there is no contribution of the SM. (a) $\phi=36.87^\circ$. (b) $\theta=160^\circ$.}
\label{dcsmpangb0}
\end{figure}

\noindent \textbf{The $(-,+)$ collision.} As already commented, there is no SM contribution to this collision. In Fig.~\ref{dcsmpangb0}, the $(-,+)$ polarized differential cross section is displayed as a function of the scattering angle and the $e_p$ angular direction $\phi$. It can be appreciated from this figure that the CESM contributions is about 2 orders of magnitude larger than the one coming from the SME and that it dominates in almost the entire variation range of the scattering angle. \textit{In this collision, effects of new physics whose source is not Lorentz violation are clearly dominant.}

\begin{figure}[htb!]
\centering
\subfigure[]{\includegraphics[scale=0.65]{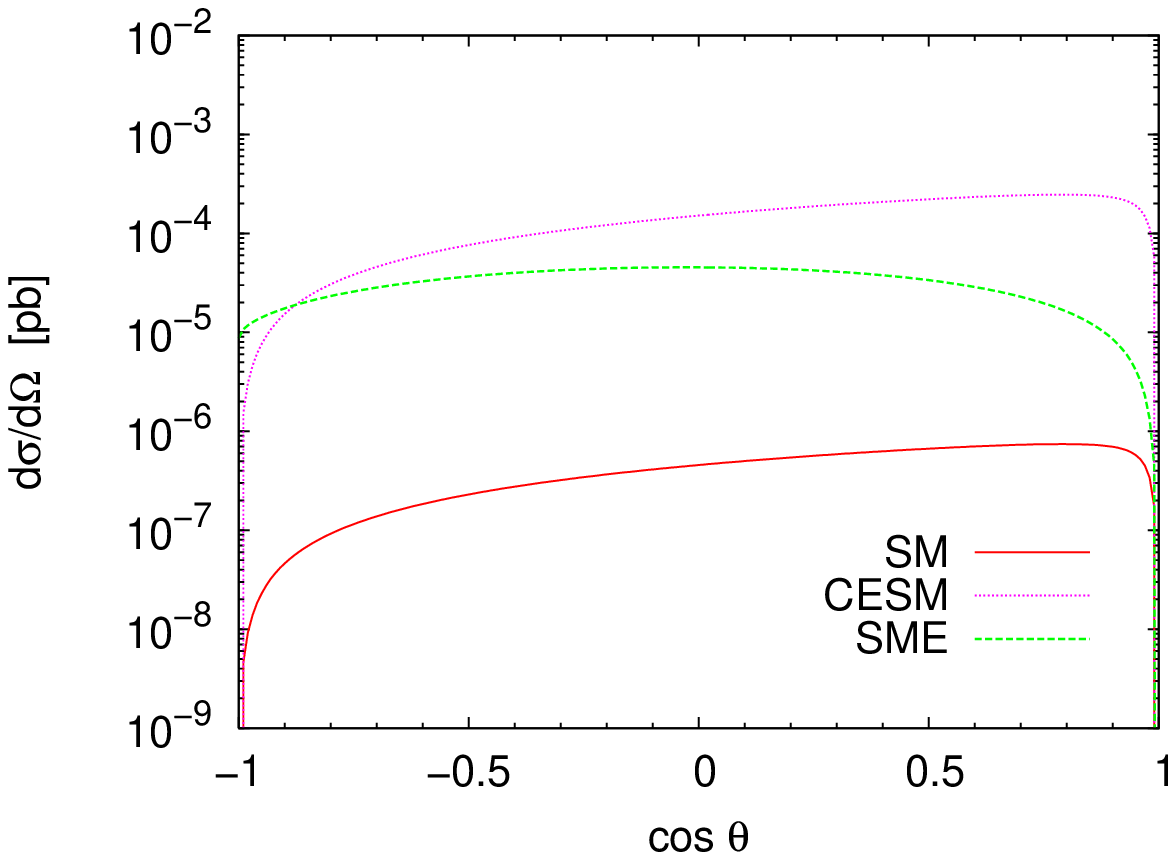}}\qquad
\subfigure[]{\includegraphics[scale=0.65]{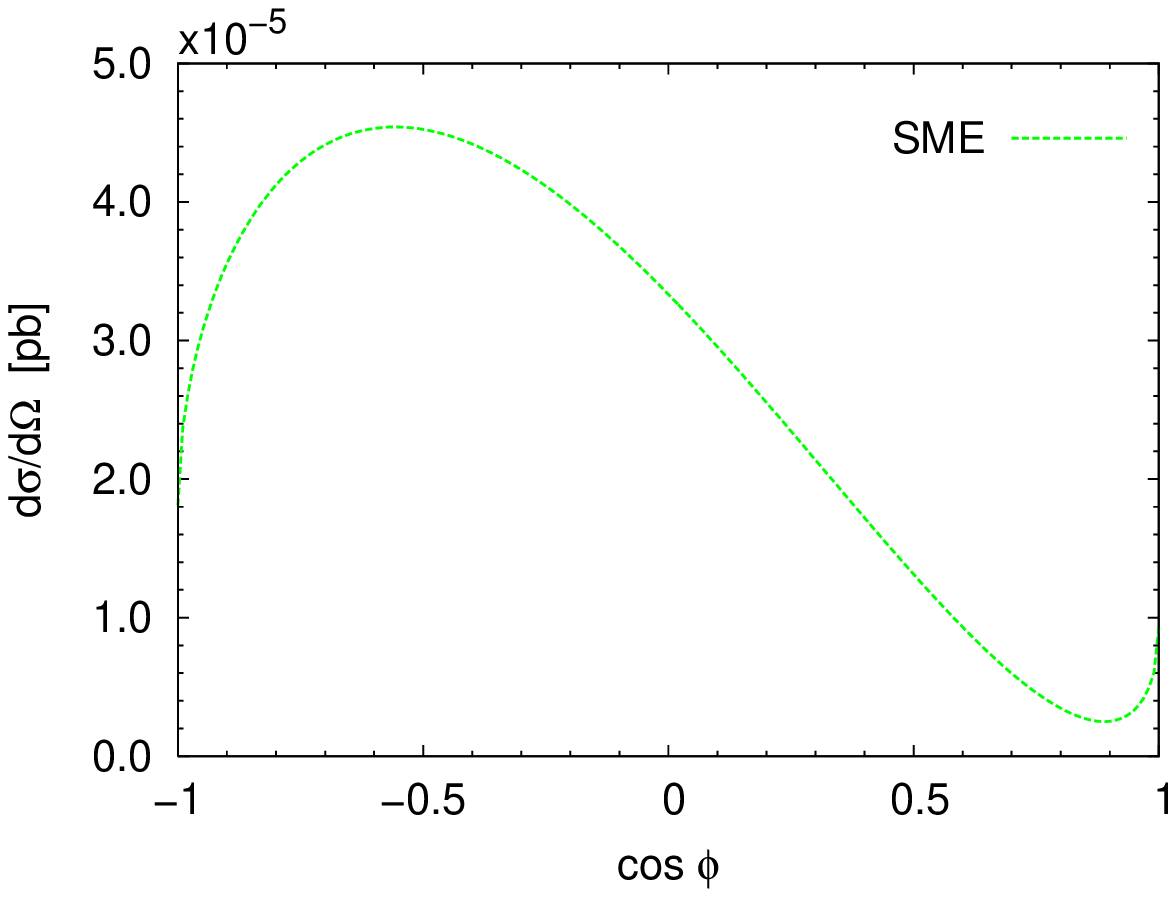}}
\caption{Differential cross section for the $e \gamma \to W\nu_e$ process with the $(+,-)$ polarization state at $\sqrt{s}=1$ TeV ($\textbf{e}\neq0$, $\textbf{b}=0$). (a) $\phi=123.37^\circ$. (b) $\theta=90^\circ$.}
\label{dcspmangb0}
\end{figure}

\noindent \textbf{The $(+,-)$ collision.} Fig.~\ref{dcspmangb0} shows the $(+,-)$ polarized differential cross section as a function of the scattering angle and the $e_p$ angular direction $\phi$. As it was seen in previous scenario, the SM contribution is subdominant in this collision. Fig.~\ref{dcspmangb0}(a) displays the differential cross section as a function of $\cos\theta$. From this figure, it can be appreciated that the CESM contribution is larger than the SME one by about one order of magnitude for $\theta= 25^\circ$. \textit{Again, new physics effects arising from the CESM dominates on those coming from the SME.}

\begin{figure}[htb!]
\centering
\subfigure[]{\includegraphics[scale=0.65]{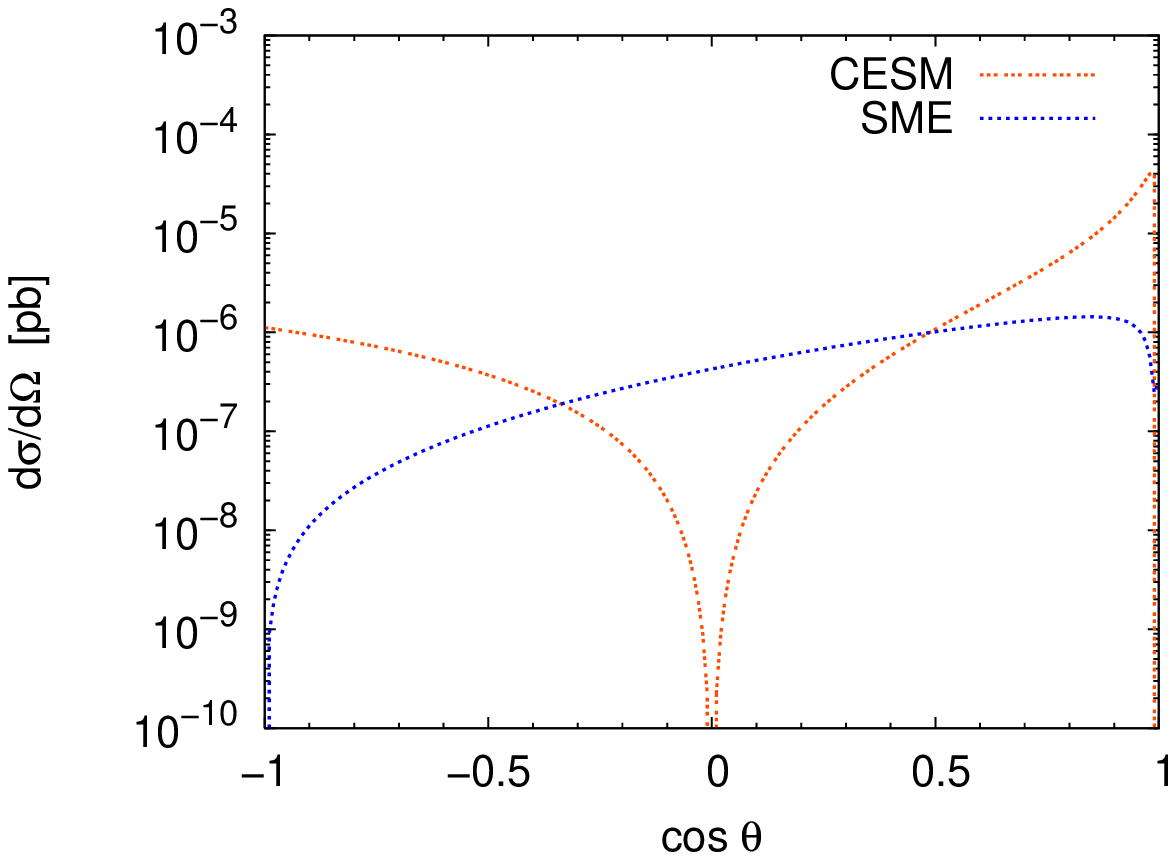}}\qquad
\subfigure[]{\includegraphics[scale=0.65]{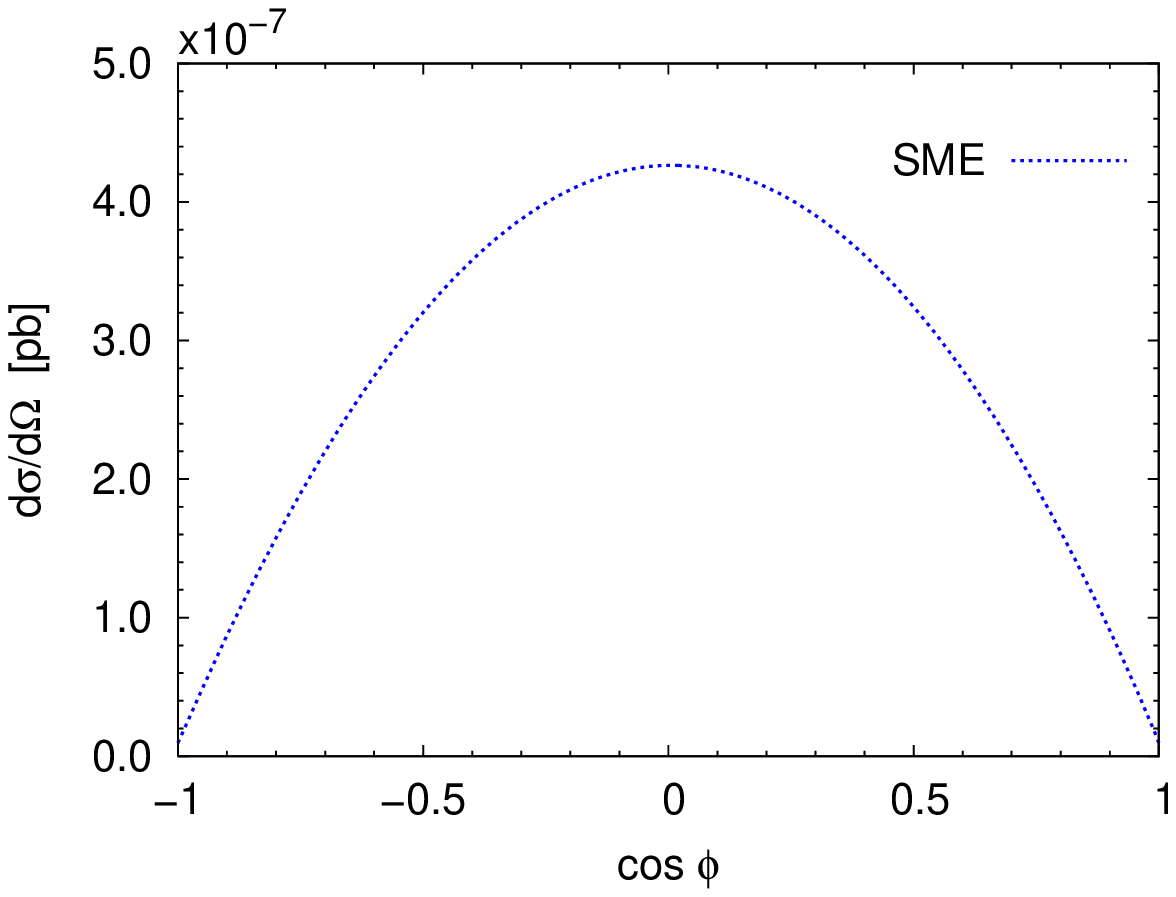}}
\caption{Differential cross section for the $e \gamma \to W\nu_e$ process with the $(-,0)$ polarization state at $\sqrt{s}=1$ TeV ($\textbf{e}\neq0$, $\textbf{b}=0$), where there is no contribution of the SM. (a) $\phi=90^\circ$. (b) $\theta=90^\circ$.}
\label{dcsmangb0}
\end{figure}

\noindent \textbf{The $(-,0)$ collision.} Recall that there is no SM contribution to this collision. In Fig.~\ref{dcsmangb0} we show the $(-,0)$ polarized differential cross section as a function of the scattering angle and the $e_p$ angular direction $\phi$. The Fig.~\ref{dcsmangb0}(a) displays the differential cross section as a function of $\cos\theta$, whereas Fig.~\ref{dcsmangb0}(b) shows that the optimal detection of Lorentz violation signal corresponds to $\phi= 90^\circ$. From both figures, Fig.~\ref{dcsmangb0}(a) and Fig.~\ref{dcsmangb0}(b), it can be appreciated a clear signal of Lorentz violation in the neighborhood of  $\theta=90^\circ$ and  $\phi =90^\circ$. \textit{This collision shows a window through which a signal of Lorentz violation can be observed.}

In summary, signals of Lorentz violations are less favored in this scenario, with the exception of the $(-,0)$ collision, which shows a clear signal in the neighborhood of  $\theta=90^\circ$ and  $\phi =90^\circ$. However, the corresponding differential cross section is about of $10^{-6}$ pb, which is 2 orders of magnitude lower than those favorable cases of Lorentz violation that emerge in the scenario with $\textbf{e}=0$, $\textbf{b}\neq 0$.

%%%%%%%%%%%%%%%%%%%%%%%%%%%%%%%%%%%%%%%%%%%%%%%%%%%%%%%%%%%%%%%%%%%%%%%%%%%%%%%%%%%%%%%

\subsubsection{Scenario $\textbf{e}\neq0$, $\textbf{b}\neq0$}
As it has been stressed from the beginning, our main interest is to find angular regions for the differential cross sections where signals of Lorentz violation can be isolated not only from the SM contribution but also from other sources of anomalous effects. In previous two scenarios analyzed, we have found that signals of Lorentz violation can be clearly observed in the scenario $\textbf{e}=0$, $\textbf{b}\neq0$, but not in the $\textbf{e}\neq 0$, $\textbf{b}=0$ one, in which effects of new physics arising from other sources dominate. Moreover, the values of the differential cross sections in the former scenario are, in general terms, 2 orders of magnitude larger than in the latter one. So, one can to expect that a more general scenario with both non-vanishing electric-like and magnetic-like  background fields does not modify essentially the prediction of the dominant scenario $\textbf{e}=0$, $\textbf{b}\neq0$. To foresee any subtle cancellation arising from some interference effect, we have performed an exhaustive analysis which showed that the prediction of the scenario $\textbf{e}=0$, $\textbf{b}\neq0$ remains essentially unchanged.

\subsection{Total cross section}
In this part of the paper, we will concentrate our attention in those angular regions in which signals of Lorentz violations show up more clearly. We will analyze  only the $\textbf{e}=0$, $\textbf{b}\neq0$ scenario which is the most promising one. As it was shown in previous subsections, signals of Lorentz violation are more clearly appreciated in the neighborhood of $\theta=160^\circ$ and  $\chi = 80^\circ$ for the differential cross sections $(+,+)$, $(+,-)$, and $(-,+)$, which are of the same order of magnitude for these parameter values. In the case of the differential cross section $(-,0)$, a clear signal of Lorentz violation is observed in the neighborhood of $\theta=90^\circ$ and  $\chi = 0^\circ$, but it is suppressed by about 2 orders of magnitude with respect to the $(+,+)$, $(+,-)$, and $(-,+)$ ones. To best appreciate the Lorentz violation effect, we will integrate the differential cross sections $(+,+)$, $(+,-)$, and $(-,+)$ in the angular interval $150^\circ <\theta <170^\circ$, with $\chi = 80^\circ$. In the case of the polarized cross section $(-,0)$, we will consider the scenario $80^\circ <\theta <100^\circ$, with $\chi = 0^\circ$.

\begin{figure}[ht]
\centering
\includegraphics[scale=0.6]{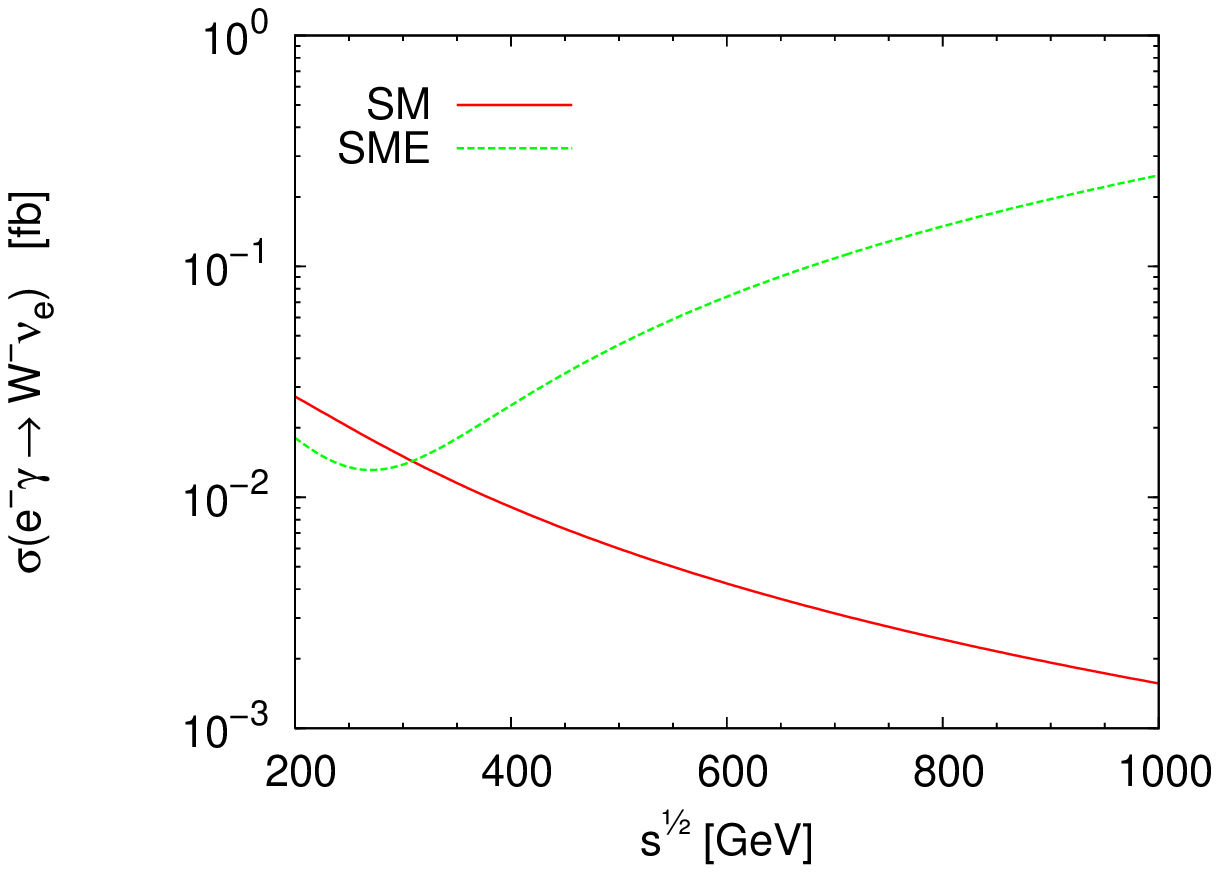}
\caption{Polarized cross section $(+,+)$ for the $e\gamma \to W\nu_e$  reaction in the interval $150^\circ <\theta <170^\circ$. Only the dominant scenario $\textbf{e}=0$, $\textbf{b}\neq0$, with $\chi=80^\circ$, is considered.}
\label{CSCPP}
\end{figure}

In Fig.~\ref{CSCPP}, the behavior of the integrated polarized cross section $(+,+)$ in the above mentioned angular window is shown as a function of the center of mass energy. From this figure, a clear effect of Lorentz violation can be appreciated starting in $\sqrt{s}\simeq$ 350 GeV, which can reach a value of up to 3 orders of magnitude above the SM signal for $\sqrt{s}\simeq$ 1000 GeV (recall that there is no CESM signal for this polarized state).

\begin{figure}[htb!]
\centering
\includegraphics[scale=0.6]{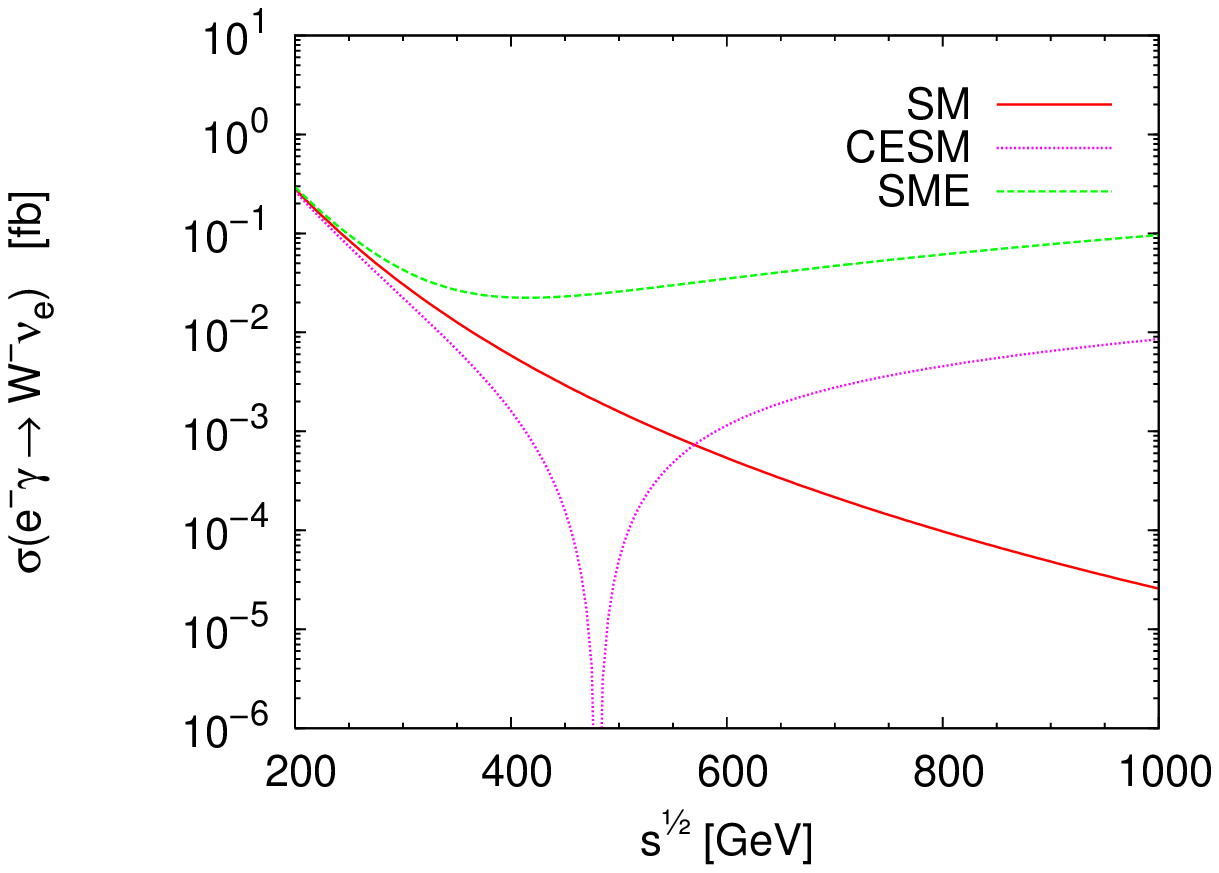}
\caption{Polarized cross section $(+,-)$ for the $e\gamma \to W\nu_e$  reaction in the interval $150^\circ <\theta <170^\circ$. Only the dominant scenario $\textbf{e}=0$, $\textbf{b}\neq0$, with $\chi=80^\circ$, is considered.}
\label{CSCPM}
\end{figure}

In Fig.~\ref{CSCPM} the behavior of the integrated polarized cross section $(+,-)$  is shown in the energy interval 200 GeV$<\sqrt{s}<$1000 GeV. From this figure, it can be observed that the SME dominates by, at least, one order of magnitude on the SM and CESM signals. As it occurs in the $(+,+)$ case, the Lorentz violation signal is significant, relative to the other signals, for energies larger than $\sqrt{s}=350$ GeV.

\begin{figure}[htb!]
\centering
\includegraphics[scale=0.6]{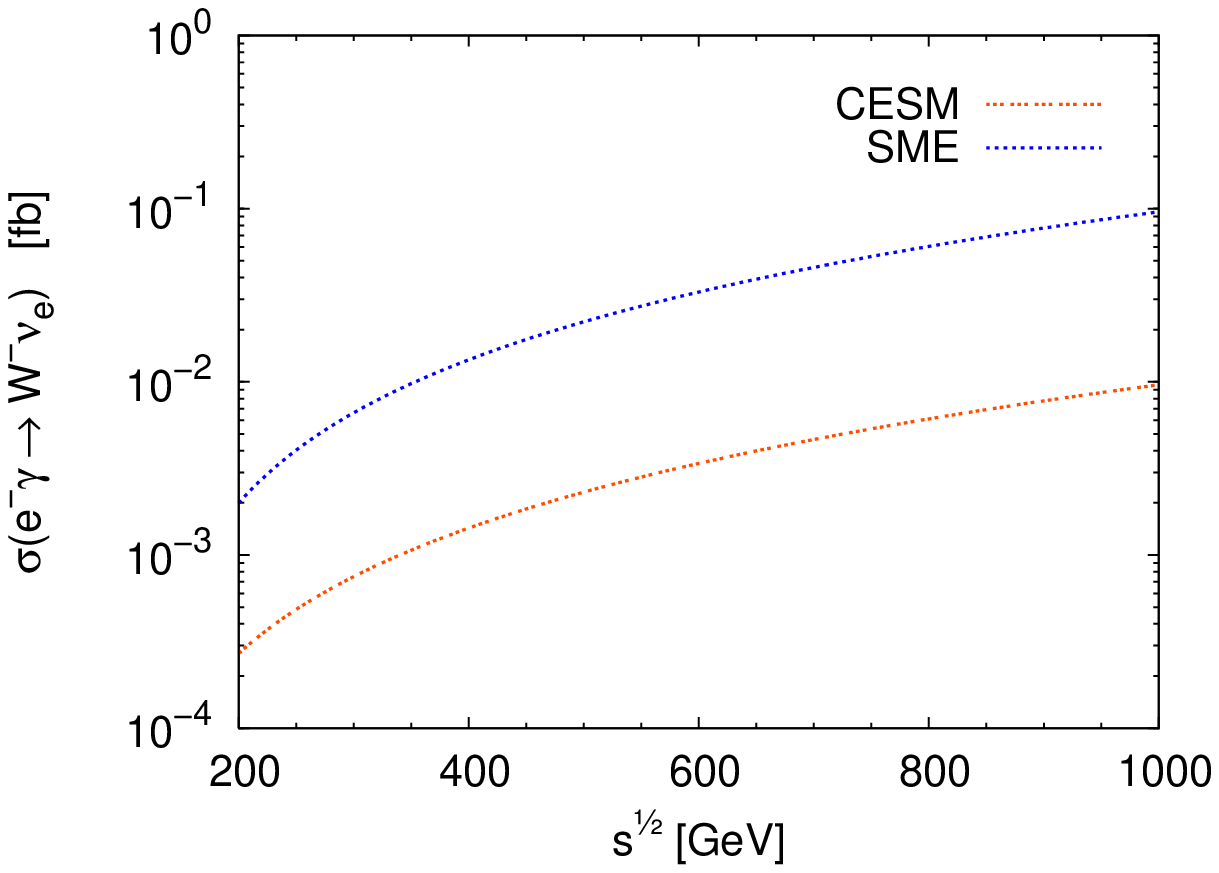}
\caption{Polarized cross section $(-,+)$ for the $e\gamma \to W\nu_e$  reaction in the interval $150^\circ <\theta <170^\circ$. Only the dominant scenario $\textbf{e}=0$, $\textbf{b}\neq0$, with $\chi=80^\circ$, is considered.}
\label{CSCMP}
\end{figure}

In Fig.~\ref{CSCMP} the integrated polarized cross section $(-,+)$ is displayed as a function of the center of mass energy. As it can be appreciated from this figure, the SME is larger than the CESM one (recall that there is no SM contribution for this state at the tree level) by about one order of magnitude for $\sqrt{s}$ ranging from $200$ GeV to $1000$ GeV.

\begin{figure}[htb!]
\centering
\includegraphics[scale=0.6]{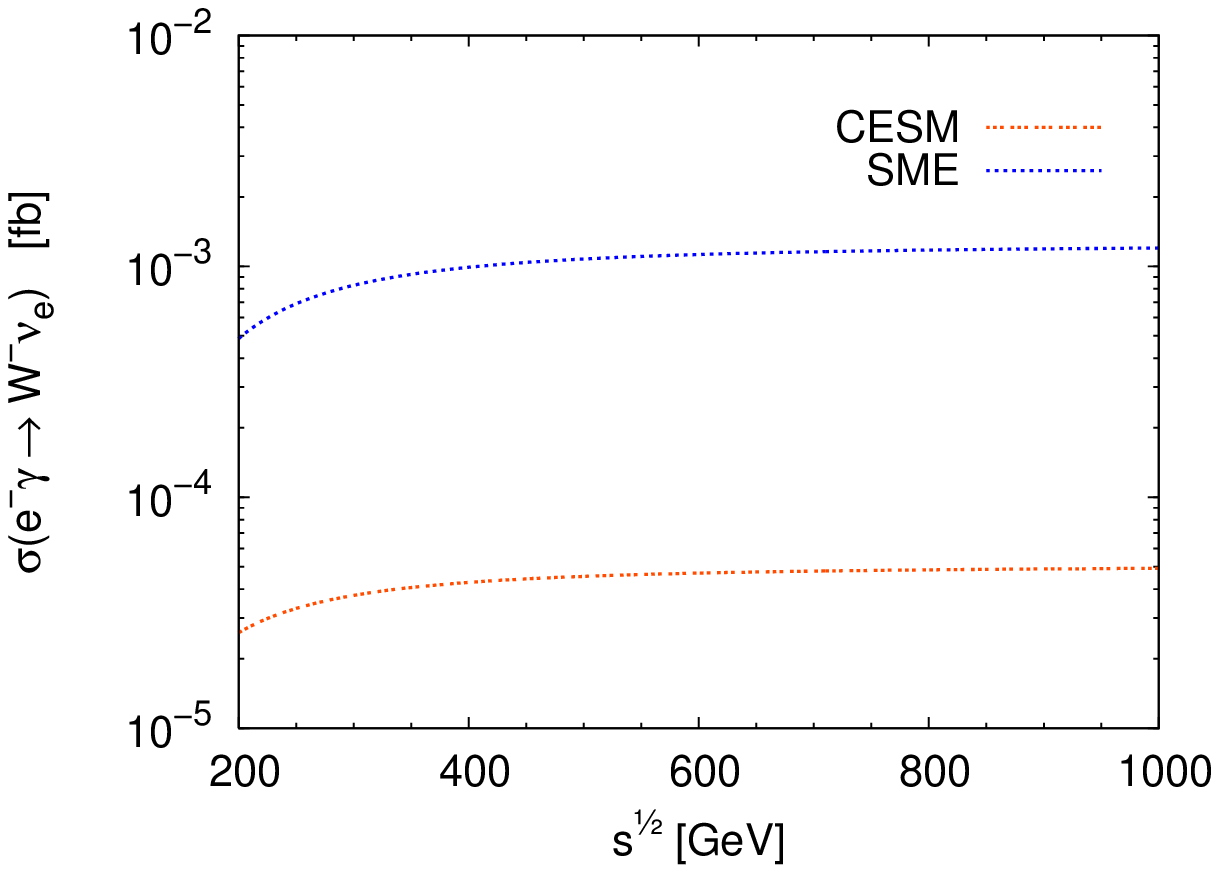}
\caption{Polarized cross section $(-,0)$ for the $e\gamma \to W\nu_e$  reaction in the interval $80^\circ <\theta <100^\circ$. Only the dominant scenario $\textbf{e}=0$, $\textbf{b}\neq0$, with $\chi=0^\circ$, is considered.}
\label{CSCM}
\end{figure}

In previous three scenarios analyzed, Lorentz violation signal arises from polarized cross sections ranging from about $10^{-2}$ fb to $10^{-1}$ fb. We now  turn to analyze the behavior of the integrated cross section for the collision $(-,0)$. Its behavior as a function of the center of mass energy is shown in Fig.~\ref{CSCM}. From this figure, it can be appreciated that the SME predicts a cross section ranging from $10^{-5}$ fb to $10^{-3}$  fb and that it is larger than that predicted by the CESM by more of one order of magnitude in all the energy interval considered. Although clear, the Lorentz violation signal is suppressed in this case by more of one order of magnitude compared with the signals that emerge in the collisions $(+,+)$, $(+,-)$, and $(-,+)$. At the ILC, it is expected an integrated luminosity of $500$ fb$^{-1}$ in the first years of operation \cite{ILC}, taking into account this we estimate around of few tens of events for a Lorentz violation signal, at best.

%%%%%%%%%%%%%%%%%%%%%%%%%%%%%%%%%%%%%%%%%%%%%%%%%%%%%%%%%%%%%%%%%%%%%%%%%%%%%%%%%%

\subsection{Asymmetry}
An observable of experimental interest is the polarization asymmetry, which is defined by~\cite{RSM}
\begin{equation}
A_{RL}=\frac{\sigma(\lambda_\gamma=+)-\sigma(\lambda_\gamma=-)}{\sigma(\lambda_\gamma=+)+\sigma(\lambda_\gamma=-)} \, .
\end{equation}
We are interested in studying the impact of Lorentz violation on this observable. As in previous subsection, we will center our analysis in the scenario in which the SME prediction is maximized, that is, $150^\circ <\theta < 170^\circ$, $\chi=80^\circ$, and $(\textbf{e}=0, b_p=1,b_y=1)$. For comparison purposes with previous works within the SM~\cite{RSM}, results without cuts in the center of mass angle ($0^\circ <\theta < 180^\circ$) will be also presented.

\begin{figure}[htb!]
\centering
\subfigure[]{\includegraphics[scale=0.65]{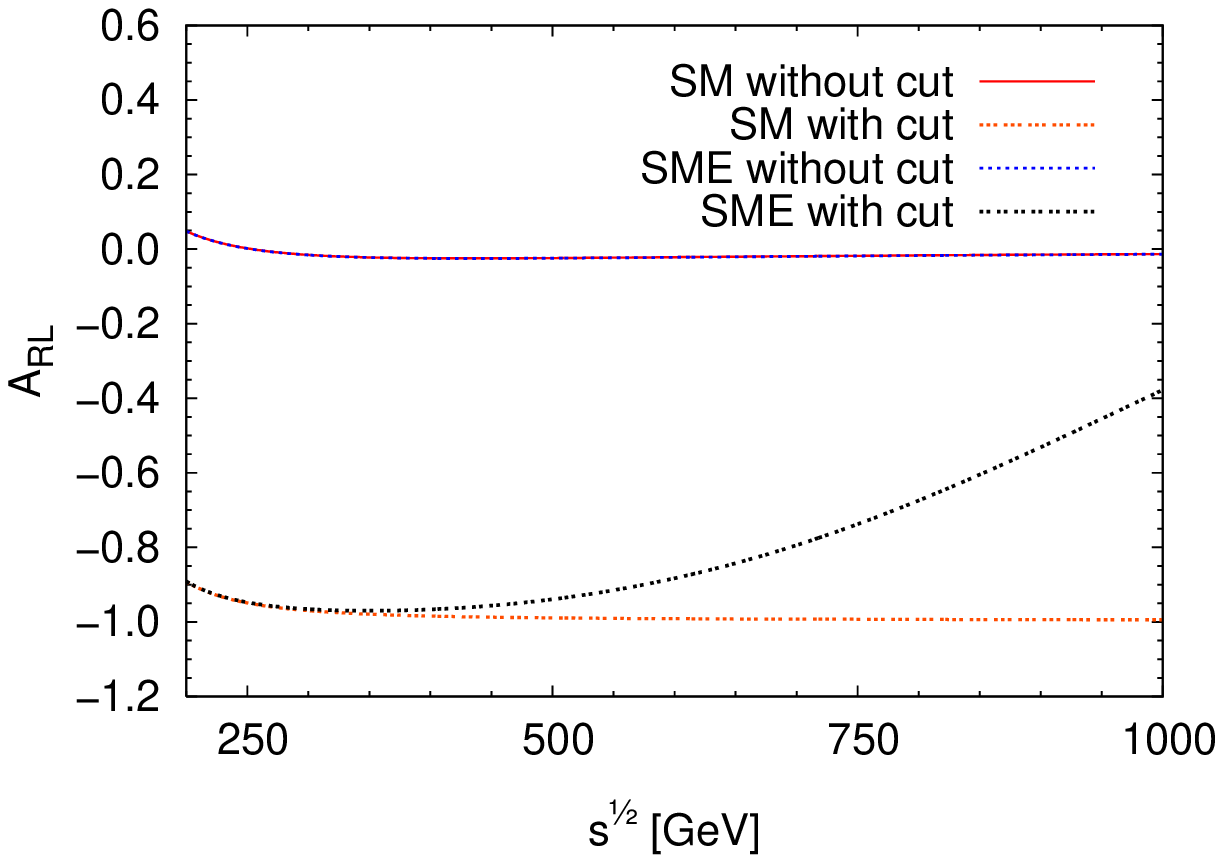}}\qquad
\subfigure[]{\includegraphics[scale=0.65]{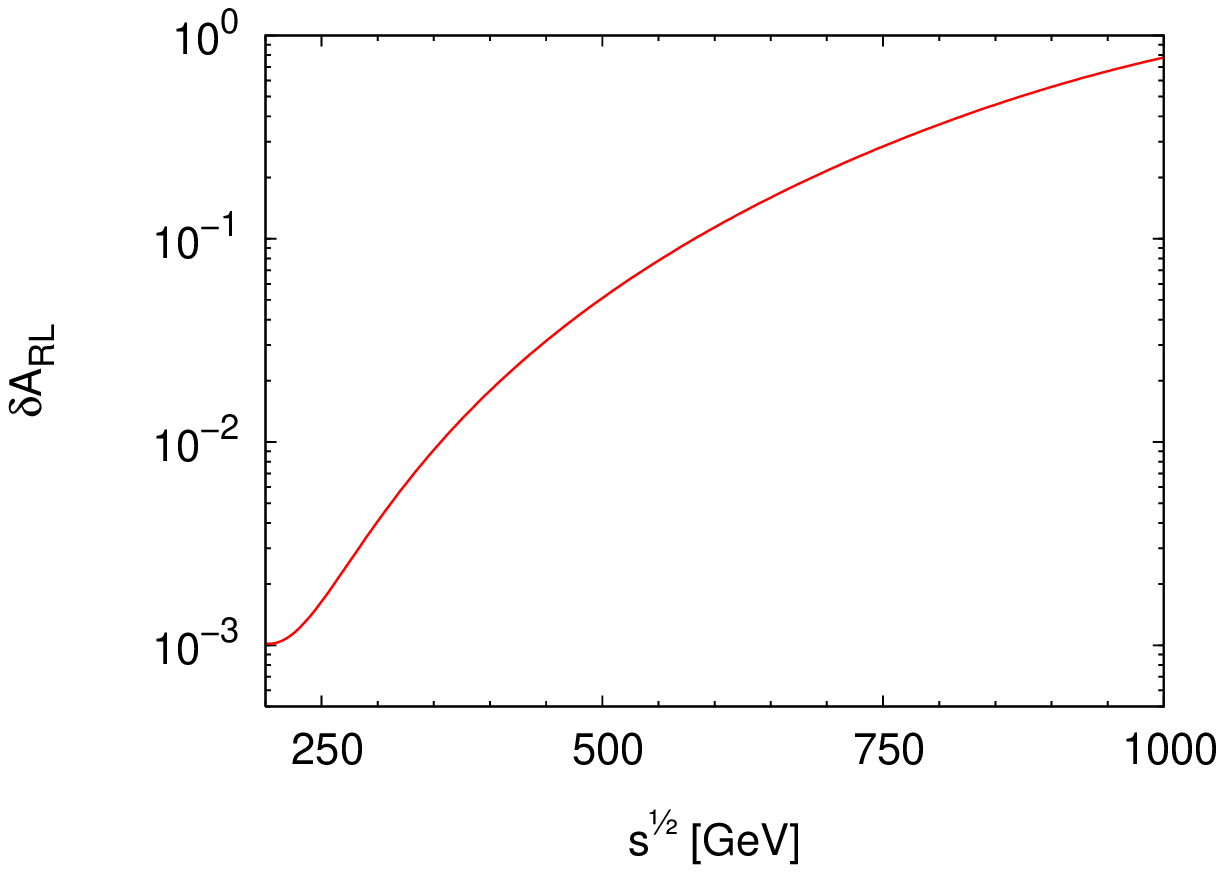}}
\caption{(a) Asymmetry for the $e\gamma \to W\nu_e$ reaction at the tree level. (b) SME correction to the tree-level SM asymmetry for the $e\gamma \to W\nu_e$ process.}
\label{ARL}
\end{figure}

In Fig.~\ref{ARL}(a), the $A_{RL}$ asymmetry is shown as a function of the center of mass energy of the collision with and without angular cut. In the first case, an angular region comprised in the interval $150^0<\theta<170^0$ was considered. We have reproduced the tree-level SM result given in Ref.~\cite{RSM}. In particular, we have verified that the SM asymmetry, $A^{SM}_{RL}$, without angular cut tends to zero at high energies. This is no the case when the angular cut is introduced. It is important to note that in this discussion we do not include the results for the CESM contribution, since this new physics effects are suppressed in the scenario with the angular cut imposed.  Fig.~\ref{ARL}(a) shows an interesting behavior for the SME asymmetry with angular cut, in which it can be appreciated that the Lorentz violation effect reduces the negative behavior of the intensity of the asymmetry with respect to the SM. Notice that this effect increases as the energy increases.

In order to amphasize the relative importance of the SME contribution to the asymmetry, we split it in the way
\begin{equation}\label{dARL}
A_{RL}=A^{SM}_{RL}+\delta A_{RL}\, ,
\end{equation}
where $A^{SM}_{RL}$ is the SM asymmetry and $\delta A_{RL}$ contains the contributions of the interference terms between SM and SME as well as the pure new physics contribution. Fig.~\ref{ARL}(b) displays $\delta A_{RL}$ as a function of the center of mass energy of the collision. At collision energies close to 1 TeV, the SME contribution to the asymmetry is above than the SM prediction. From this figure, it can be observed that the pure new physics effect of the SME leads to a deviation in the asymmetry quite close to the unit for $\sqrt{s}=1$ TeV, which constitutes a clear signal of Lorentz violation.

\section{Conclusions}
\label{C} Effects of Lorentz violation could be detected in future high energy experiments. This class of new physics effects can induce deviations on observables that are sensitive to spatial orientation, as the presence of preferred spacetime directions, which is the main feature of Lorentz violation. These type of effects could be insignificant if the observable in consideration is overage on all spatial directions, but they could show up with dominant intensity in some preferred directions. This is the case of polarized cross sections associated with collision processes involving particles with nonzero spin, which usually are strongly depending on the scattering angle.

In this paper, the center-mass helicity amplitudes for the $\gamma e\to W\nu_e$ process where derived in the context of both the SME and the CESM. The former is a gauge invariant SM extension which incorporates Lorentz violation in a model-independent way, whereas the latter one is an effective theory which incorporates new physics effects, also in a model-independent fashion, but subject to respect both the Lorentz and the SM gauge symmetries. An effective Yang-Mills Lagrangian for the electroweak sector of the SM was constructed through the introduction of a dimension-six Lorentz 2-tensor ${\cal O}_{\alpha \beta}$ operator, which is invariant under the $SU_L(2)$ group. New physics effects in the context of CESM were incorporated through the Lorentz invariant $g^{\alpha \beta}{\cal O}_{\alpha \beta}$. On the other hand, nonrenormalizable effects of Lorentz violation were considered in the context of the SME via the observer invariant $b^{\alpha \beta}{\cal O}_{\alpha \beta}$, with $b^{\alpha \beta}$ an antisymmetric constant 2-tensor. The six components of this constant tensor were parametrized in terms of the electric-like $\textbf{e}$ and magnetic-like $\textbf{b}$ spatial vectors. It was found that the best scenario in which the signal of Lorentz violation is larger than the background (SM and CESM effects) corresponds to $\textbf{e}=0$ and $\textbf{b}\neq 0$, in the the neighborhood of $\theta=160^\circ$ and  $\chi = 80^\circ$, where $\theta$ and $\chi$ are the scattering angle and the angle formed by the component of $\textbf{b}$ parallel to the collision plane, respectively. The signal arise essentially from the polarized cross sections $(+,+)$, $(+,-)$, and $(-,+)$. For $|\textbf{b}|\sim 1$ and $\chi = 80^\circ$, the integration of these polarized differential cross section on the $150^\circ<\theta<170^\circ$ interval leads to values which are between one and two orders of magnitude larger than the background effects. The signal of Lorentz violation ranges from $10^{-2}$ fb to $10^{-1}$ fb, at best. With the integrated luminosity of $500$ fb$^{-1}$ expected in the ILC during the first years of operation, up to tens of events should be observed.

\acknowledgments{We acknowledge financial support from CONACYT, CIC-UMSNH and
SNI (M\' exico).}

\end{document}